\documentclass{article}
\usepackage[utf8]{inputenc}
\usepackage[T1]{fontenc}
\usepackage{amsmath}
\usepackage{amsthm,thm-restate}
\usepackage{amssymb}
\usepackage{fullpage}
\usepackage[disableredefinitions]{complexity}
\usepackage{framed}
\usepackage{bold-extra}
\usepackage{graphicx}
\usepackage{xspace}
\usepackage[auth-sc]{authblk}
\usepackage{authblk}

\usepackage{hyperref}

\usepackage[linesnumbered,ruled,vlined]{algorithm2e}

\usepackage{booktabs}

\usepackage{tikz}
\usetikzlibrary{positioning}
\usetikzlibrary{shapes}
\usetikzlibrary{fit}
\tikzstyle{vert} = [circle,fill=black, minimum size=2mm, inner sep=0pt]
\tikzstyle{edge} = [thick]

\usepackage{tkz-euclide}
\usepackage{xstring}

\newtheorem{theorem}{Theorem}[section]

\newtheorem{lemma}[theorem]{Lemma}
\newtheorem{example}[theorem]{Example}
\newtheorem{definition}[theorem]{Definition}

\newtheorem{cor}[theorem]{Corollary}
\DeclareMathOperator{\ND}{ND}

\newcommand{\StatOpt}{\textsc{Correlation Subgraph Optimisation}\xspace}
\newcommand{\countkStatOpt}{\textsc{Enum}-$k$-\textsc{Correlation Subgraph Optimisation}\xspace}
\newcommand{\kStatOpt}{$k$-\textsc{Correlation Subgraph Optimisation}\xspace}
\newcommand{\AvNbdOpt}{\textsc{Average Value Neighbourhood Optimisation}}

\newcommand{\fancySAT}{\textsc{Cubic Planar Monotone 1-in-3 SAT}}
\newcommand{\nideal}{\textsc{Near Ideal Subgraph}\xspace}
\newcommand{\ssum}{\textsc{Subset Sum}\xspace}

\newcommand{\enumEmbedCycle}{\textsc{Enumerate Embed Coloured Cycle}\xspace}

\DeclareMathOperator{\true}{TRUE}
\DeclareMathOperator{\false}{FALSE}

\DeclareMathOperator{\cont}{cont}

\DeclareMathOperator{\score}{score}

\usepackage{orcidlink}

\title{The complexity of finding and enumerating optimal subgraphs to represent spatial correlation}
\author[1]{Jessica Enright~\orcidlink{0000-0002-0266-3292}}
\author[2]{Duncan Lee~\orcidlink{0000-0002-6175-6800}}
\author[1]{Kitty Meeks~\orcidlink{0000-0001-5299-3073}}
\author[1]{William Pettersson\footnote{Corresponding author}~\orcidlink{0000-0003-0040-2088}}
\author[1]{John Sylvester~\orcidlink{0000-0002-6543-2934}}

\affil[1]{School of Computing Science, University of Glasgow}
\affil[2]{School of Mathematics and Statistics, University of Glasgow}

\affil{Emails: \texttt{\{firstname.lastname\}@glasgow.ac.uk}}

\begin{document}
\maketitle

\begin{abstract}
Understanding spatial correlation is vital in many fields including epidemiology and social science.
Lee, Meeks and Pettersson (Stat.\ Comput.\ 2021) recently demonstrated that improved inference for areal unit count data can be achieved by carrying out modifications to a graph representing spatial correlations; specifically, they delete edges of the planar graph derived from border-sharing between geographic regions in order to maximise a specific objective function.  In this paper we address the computational complexity of the associated graph optimisation problem.
We demonstrate that this problem cannot be solved in polynomial time unless P = NP; we further show intractability for two simpler variants of the problem.
We follow these results with two parameterised algorithms that exactly solve the problem.
Both of these solve not only the decision problem, but also enumerate all solutions with polynomial time precalculation, delay, and postcalculation time in respective restricted settings.
For this problem, efficient enumeration allows the uncertainty in the spatial correlation to be utilised in the modelling.
The first enumeration algorithm utilises dynamic programming on a tree decomposition, and has polynomial time precalculation and linear delay if both the treewidth and maximum degree are bounded.
The second algorithm is restricted to problem instances with maximum degree three, as may arise from triangulations of planar surfaces, but can output all solutions with FPT precalculation time and linear delay when the maximum number of edges that can be removed is taken as the parameter.
\end{abstract}

%\keywords{Parameterised complexity, treewidth, colour coding, spatial statistics}

\section{Introduction}
Spatio-temporal count data relating to a set of $n$ non-overlapping areal units for $T$ consecutive time periods are prevalent in many fields, including epidemiology \cite{stoner2019} and social science \cite{bradley2016}.
As geographical proximity can often indicate correlation, 
such data can be modelled as a graph, with vertices representing areas and edges between areas that share a geographic boundary and so their data values are assumed to be spatially correlated.
This spatial correlation is then represented as a weight matrix arising from these adjacency relationships.
However, such models are often not ideal representations as geographical proximity does not always imply correlation~\cite{mitchell2014}.
Instead,
Lee, Meeks and Pettersson~\cite{statspaper} recently proposed a new method for addressing this issue by deriving a specific objective function (given in full in Section~\ref{sec:statopt}), and then searching for a spanning subgraph with no isolated vertices which maximises this function.
Maximising this objective function corresponds to maximising the natural log of the product of full conditional distributions over all vertices (corresponding to spatial units) in a conditional autoregressive model; further details are available in~\cite{statspaper}.
This objective function is highly non-linear, and rewards removing as few edges as possible, while applying a penalty that (non-linearly) increases as the difference between the weight of each vertex and the average weight over its neighbours increases.
Lee, Meeks and Pettersson~\cite{statspaper} studied an application of this problem where the geographical proximity data corresponds to a graph with 257 vertices; we note that an exhaustive search for an optimal subgraph would be intractable even with significantly fewer vertices. 
Instead, Lee, Meeks and Pettersson gave a heuristic for solving this problem, but point out that many standard techniques are not applicable to this problem, suggesting that it is challenging to efficiently find even one optimal subgraphs.

In addition, 
the overarching approach when modelling spatial correlation in areal unit data is to fix the graph that corresponds to a spatial neighbourhood matrix in advance of undertaking the modelling. 
However, the neighbourhood matrix (based on the graph) determines the spatial correlation structure in the data, and hence should be estimated as a parameter in the model.
Most approaches either ignore this estimation problem or estimate a single value for it (i.e.,~assume it can be approximated by geographical contiguity), but both ignore the uncertainty in its value.
This can be improved by using enumeration algorithms (i.e.~algorithms that find all optimal subgraphs rather than just one) to model
this uncertainty, which can be incorporated via a Bayesian modelling framework~\cite{Lee2014}.

\subsection{Our contribution}

We show that the problem is indeed NP-hard, even on planar graphs, 
and provide examples that illustrate two of the major challenges inherent in the problem: we cannot optimise independently on disjoint connected components and we cannot iterate towards a solution. 
We also show that the decision variant of minimising the penalty portion of the objective function is NP-complete even when restricted to planar graphs with maximum degree at most five.
We then investigate a simplification in which the goal is to find a subgraph with a penalty term of zero.  We completely characterise all such subgraphs, and then show that the problem is solvable in linear time in the number of edges of the graph.
However, we also show that finding a subgraph with a penalty term of zero on all vertices of degree two or more is NP-complete.

In the positive direction, we give two exact algorithms that are tractable in respective restricted settings, and enumerate all optimal solutions.
Both algorithms have connections to the maximum degree of the input graph: we note that graphs arising from areal studies will often have small maximum degree.
The first algorithm has a precalculation time of $f(tw(G),\Delta(G)) \cdot n^{O(\Delta(G))}$, $O(n^2)$ delay, and $O(n^2)$ postcalculation time, where $tw(G)$ and $\Delta(G)$ are the treewidth and maximum degree of $G$ respectively.
The second algorithm is only guaranteed to be correct if the underlying graph has maximum degree three: the consideration of such restrictions has been increasingly common since Lindgren et al.~\cite{Lindgren2011} developed their stochastic partial differential equations approach based on a triangulation of the spatial study region.
This second algorithm enumerates all optimal subgraphs with a precalculcation time of $f(k) \cdot n \log n$ and a linear delay between outputs, where $k$ is by the maximum number of edges that can be removed.

\subsection{Paper outline}
Section~\ref{sec:background} gives notation and definitions, the formal problem definition, and examples that illustrate two of the major challenges inherent in the problem.
We then prove in Section~\ref{sec:hardness} that, unless P = NP, there is no polynomial-time algorithm to solve the main optimisation problem, even when restricted to planar graphs.
Section~\ref{sec:simple} then examines three simplifications of the problem.
In Section~\ref{sec:algs} we introduce two algorithms to exactly solve the problem, and we finish with concluding thoughts and open problems in Section~\ref{sec:conclusion}.

\section{Background}\label{sec:background}
In this section we give the notation we need for this paper, define the problem, and then demonstrate why some common techniques from graph theory are not applicable to this problem.

\subsection{Notation and definitions}
\label{sec:notation}

A graph is a pair $G=(V,E)$, where the \emph{vertex set} $V$ is a finite set, and the \emph{edge set} $E \subseteq V^{(2)}$ is a set of unordered pairs of elements of $V$.  Two vertices $u$ and $v$ are said to be \emph{adjacent} if $e = uv \in E$; $u$ and $v$ are said to be the \emph{endpoints} of $e$.  The \emph{neighbourhood} of $v$ in $G$ is the set $N_G(v) := \{u \in V: uv \in E\}$, and the \emph{degree} of $v$ in $G$ is $d_G(v) := |N_G(v)|$.  An \emph{isolated vertex} is a vertex of degree zero, and a \emph{leaf} is a vertex of degree one. The \emph{maximum degree} of a graph $G$ is $\Delta(G) := \max_{v\in V} d_G(v)$.
A graph $H = ( V_H,E_H)$ is a \emph{subgraph} of $G$ if $V_H \subseteq V$ and $E_H \subseteq E$; $H$ is a \emph{spanning subgraph} of $G$ if $V_H = V$ so that $H$ is obtained from $G$ by deleting a (possibly empty) subset of edges. Given an edge $e$ in $E(G)$ (respectively a set $E' \subseteq E(G)$) we write $G \setminus e$ (respectively $G \setminus E'$) for the subgraph of $G$ obtained by deleting $e$ (respectively deleting every element of $E'$).

A graph $G$ is \emph{planar} if it can be drawn in the plane (i.e. vertices can be mapped to points in the plane, and edges to curves in the plane whose extreme points are the images of its endpoints)  in such a way that no two edges cross.  Given any partition of a subset of the plane into regions, we can define a planar graph whose vertices are in bijection with the set of regions, in which two regions are adjacent if and only if they share a border of positive length.
In particular, if each region has three sides (i.e,~the partition is a triangulation of a subset of the plane) then the resulting graph will have maximum degree three.

\subsection{The optimisation problem}
\label{sec:statopt}
Following Lee, Meeks and Pettersson \cite{statspaper}, we are concerned with the following optimisation problem.

\begin{framed}
\noindent
\textbf{\StatOpt} \\
\textit{Input:} A graph $G = (V,E)$ where $|V| = n$, and function $f: V \rightarrow \mathbb{Q}$.\\
\textit{Question:} What is the maximum value of 
$$ \score(H,f):= \sum_{v \in V}\ln d_H(v) - n \ln \left[ \sum_{v \in V} d_H(v) \left(f(v) - \frac{\sum_{u \in N_H(v)} f(u)}{d_H(v)}\right)^2\right],$$
taken over all spanning subgraphs $H$ of $G$ such that $d_H(v) \ge 1$ for all $v \in V$?
\end{framed}

We will say that a subgraph $H$ of $G$ is \emph{valid} if $H$ is a spanning subgraph of $G$ and $d_H(v) \ge 1$ for all $v \in V$.
Given a vertex $v$ in the input graph $G$, we will sometimes refer to $f(v)$ as the \emph{weight} of $v$.
We also define the neighbourhood discrepancy of a vertex $f$ in a graph $H$ with weight function $f$ (written $\ND_H(v,f)$) as 
\[
\ND_H(v,f) := \left( f(v) - \frac{\sum_{u \in N_H(v)} f(u)}{d_H(v)}\right)^2.
\]

\subsection{Why common graph algorithm techniques fail}
\label{sec:nonlocal}
This problem is particularly resistant to many approaches common in algorithmic graph theory.
We will describe two of these now.
Firstly, on a disconnected graph $G$, combining optimal solutions on each connected component is not guaranteed to find an optimal solution on $G$.
This is true even if there are only two disconnected components, one of which is an isolated edge and the other being a path, as illustrated in the following example.

\begin{example}\label{ex:path-and-edge}
Consider the graph $G$ consisting of a path on four vertices $(v_1,v_2,v_3,\allowbreak v_4)$ along with an isolated edge between vertices $v_a$ and $v_b$, as shown in Figure~\ref{fig:ex:path-and-edge}, and let
$H = G\setminus \{v_2v_3\}$.
Note that $H$ is the only proper subgraph of $G$ which has no isolated vertices.
Let $f$ be defined as follows: $f(v_1) = 0$, $f(v_2) = 1$, $f(v_3) = 10$, $f(v_4) = 11$, $f(v_a) = 0$, and $f(v_b) = x$ for some real $x$.
If $x=1$ then $\score(G,f) < \score(H,f)$ but if $x=1000$ then $\score(G,f) > \score(H,f)$.
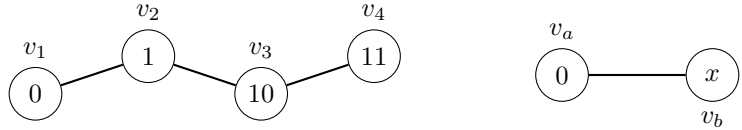
\begin{figure}
\centering
\begin{tikzpicture}
\tikzstyle{vert} = [circle,draw, minimum size=7mm, inner sep=0pt]
\node[vert,label=above:{$v_1$}] (one) at (0,0) {$0$};
\node[vert,label=above:{$v_2$}] (two) at (1.5,0.5) {$1$};
\node[vert,label=above:{$v_3$}] (three) at (3,0) {$10$};
\node[vert,label=above:{$v_4$}] (four) at (4.5, 0.5) {$11$};
\node[vert,label=above:{$v_a$}] (a) at (7, 0.25) {$0$};
\node[vert,label=below:{$v_b$}] (b) at (9, 0.25) {$x$};
\draw[edge] (one) -- (two) -- (three) -- (four);
\draw[edge] (a) -- (b);
\end{tikzpicture}
\caption{Graph for Example~\ref{ex:path-and-edge}. The value of the function at each vertex is shown inside the respective vertex.}\label{fig:ex:path-and-edge}
\end{figure}
\end{example}
To understand why disconnected components can affect each other in such a manner, note that the negative term in the score function contains a logarithm of a sum of neighbourhood discrepancies.
This means that the relative importance of the neighbourhood discrepancies of any set of vertices depends on the total sum of the neighbourhood discrepancies across the whole graph.
In other words, the presence of a large neighbourhood discrepancy elsewhere (even in a separate component) in the graph can reduce the impact of the neighbourhood discrepancy at a given vertex or set of vertices.
However, the positive term in the score function is a sum of logarithms, so the contribution to the positive term from the degree of one vertex does not depend on any other part of the graph.

A reader might also be tempted to tackle this problem by identifying a ``best'' edge to remove and proceeding iteratively. 
The following example highlights that any algorithm using such a greedy approach may, in some cases, not find an optimal solution.

\begin{example}\label{ex:path}
Consider the graph $G$ being a path on six vertices labelled $v_1, v_2, v_3,\allowbreak v_4, v_5$, and $v_6$ with $f(v_1) = 1000$, $f(v_2) = 2000$, $f(v_3) = 1999$, $f(v_4) = 1001$, $f(v_5) = 2019$, and $f(v_6) = 981$ as shown in Figure~\ref{fig:ex:path}.
Let $H = G\setminus \{v_2v_3, v_4v_5\}$,
and let $H' = G\setminus \{v_3v_4\}$.
The maximum score that can be achieved with the removal of only one edge is achieved by removing edge $v_3v_4$ and creating $H'$.
However, the optimal solution to \StatOpt on $G$ is $H$, and involves removing edges $v_2v_3$ and $v_4v_5$.
\begin{figure}
\centering
\begin{tikzpicture}
\tikzstyle{vert} = [circle,draw, minimum size=9mm, inner sep=0pt]
\node[vert,label=above:{$v_1$}] (one) at (0,0) {$1000$};
\node[vert,label=above:{$v_2$}] (two) at (2,0.5) {$2000$};
\node[vert,label=above:{$v_3$}] (three) at (4,0) {$1999$};
\node[vert,label=above:{$v_4$}] (four) at (6, 0.5) {$1001$};
\node[vert,label=above:{$v_5$}] (a) at (8, 0) {$2019$};
\node[vert,label=above:{$v_6$}] (b) at (10, 0.5) {$981$};
\draw[edge] (one) -- (two) -- (three) -- (four) -- (a) -- (b);
\end{tikzpicture}
\caption{Graph for Example~\ref{ex:path}. The value of the function at each vertex is shown inside the respective vertex.}\label{fig:ex:path}
\end{figure}
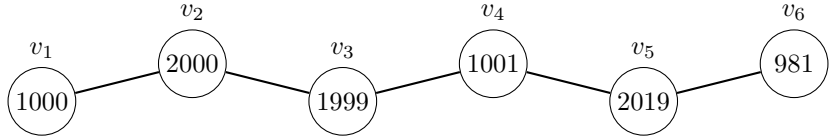
\end{example}

\section{Hardness on planar graphs}
\label{sec:hardness}

In this section we prove NP-hardness of \StatOpt on planar graphs.

\begin{restatable}{theorem}{theoremnphard}\label{thm:hard}
There is no polynomial-time algorithm to solve \StatOpt on planar graphs unless P = NP.
\end{restatable}

We prove this result by means of a reduction from the following problem, shown to be NP-complete in \cite{moore01}; the \emph{incidence graph} $G_{\Phi}$ of a CNF formula $\Phi$ is a bipartite graph whose vertex sets correspond to the variables and clauses of $\Phi$ respectively, and in which a variable $x$ and clause $C$ are connected by an edge if and only if $x$ appears in $C$.

\begin{framed}
\noindent
\textbf{\fancySAT}\\
\textit{Input:} A 3-CNF formula $\Phi$ in which every variable appears in exactly three clauses, variables only appear positively, and the incidence graph $G_{\Phi}$ is planar.\\
\textit{Question:} Is there a truth assignment to the variables of $\Phi$ so that exactly one variable in every clause evaluates to $\true$?
\end{framed}

We begin by describing the construction of a graph $G$ and function $f:V(G) \rightarrow \mathbb{N}$ corresponding to the formula $\Phi$ in an instance of \fancySAT; the construction will be defined in terms of an integer parameter $t \geq 1$ whose value we will determine later.
Note that $G$ is not the incidence graph $G_\Phi$ of $\Phi$, but instead $G_\Phi$ is a graph minor of $G$.

Suppose that $\Phi$ has variables $x_1,\ldots,x_n$ and clauses $C_1,\ldots,C_m$.  Since every variable appears in exactly three clauses and each clause contains exactly three variables, we must have $m=n$.

For each variable $x_i$, $G$ contains a variable gadget on $3t^2 + 6t + 8$ vertices.  The non-leaf vertices of the gadget are:
\begin{itemize}
\item $u_i$, with $f(u_i) = 7t$,
\item $v_i$, with $f(v_i) = 4t$,
\item $z_i$ with $f(z_i) = t$,
\item $z_i'$ with $f(z_i') = 4t$, and
\item $w_{i,j}$ for each $j \in \{1,2,3\}$, with $f(w_{i,j}) = 3t$.
\end{itemize} 
The vertex $v_i$ is adjacent to $u_i$, $z_i$ and each $w_{i,j}$ with $i \in \{1,2,3\}$; $z_i$ is adjacent to $z_i'$.  We add leaves to this gadget as follows:
\begin{itemize}
\item $u_i$ has $3t$ pendant leaves, each assigned value $7t+1$ by $f$;
\item $z_i$ has $3t$ pendant leaves, each assigned value $t-1$ by $f$;
\item $z_i'$ has $3t^2$ pendant leaves, each assigned value $4t$ by $f$;
\item each vertex $w_{i,j}$ has exactly one pendant leaf, assigned value $3t$ by $f$.
\end{itemize}

For each clause $C_j$, $G$ contains a clause gadget on $t^2+2$ vertices: $a_j$ and $a_j'$, which are adjacent, and $t^2$ pendant leaves adjacent to $a_j'$.  We set $f(a_j)=2t$, and $f$ takes value $t$ on $a_j'$ and all of its leaf neighbours.

\begin{figure}[htb]
\centering
\includegraphics[width = \linewidth]{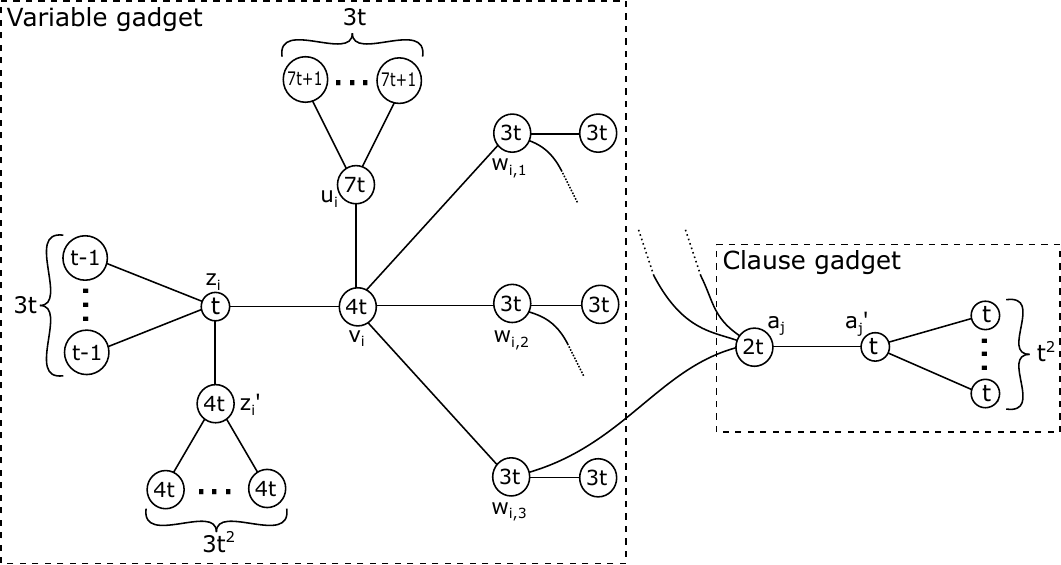}
\caption{Construction of the variable and clause gadgets.}\label{fig:gadgets}
\end{figure}

We complete the definition of $G$ by specifying the edges with one endpoint in a variable gadget and the other in a clause gadget: if the variable $x_i$ appears in clauses $C_{r_1}$, $C_{r_2}$ and $C_{r_3}$, with $r_1 < r_2 < r_3$, then we have edges $w_{i,1}a_{r_1}$, $w_{i,2}a_{r_2}$ and $w_{i,3}a_{r,3}$. The construction of the variable and clause gadgets is illustrated in Figure \ref{fig:gadgets}.

Recall that a subgraph $H$ of $G$ is \emph{valid} if $H$ is a spanning subgraph of $G$ and $d_H(v) \ge 1$ for all $v \in V$.
Recall that the \emph{neighbourhood discrepancy} of a vertex $v$ with respect to $f$ in a valid subgraph $H$, written $\ND_H(v,f)$, is
$$\ND_H(v,f) := \left(f(v) - \frac{\sum_{u \in N_H(v)} f(v)}{d_H(v)}\right)^2.$$ 
The goal of \StatOpt is therefore to maximise
$$\score(H,f):= \sum_{v \in V} \ln d_H(v) - n \ln \left[\sum_{v \in V}d_H(v) \ND_H(f,v)\right],$$
over all valid subgraphs $H$ of $G$.  

We now prove several properties of valid subgraphs of $G$.

\begin{lemma}\label{clm:ND-leaves}
For any valid subgraph $H$, 
$$\sum_{u \text{ a leaf in } G} \ND_H(u,f) = 6nt.$$
\end{lemma}
\begin{proof}
Note that the neighbourhood of any leaf vertex must be the same in $G$ and in any valid subgraph $H$; it therefore suffices to determine $\sum_{u \text{ a leaf in } G} \ND_G(u,f)$.  In each variable gadget, there are $3t^2 + 3$ leaves whose neighbourhood discrepancy in $G$ is zero, and $6t$ leaves for which the neighbourhood discrepancy is one.  This gives a contribution to the sum of $6t$ for every variable gadget, a total of $6nt$ over all such gadgets.  In each clause gadget there are $t^2$ leaves, each with neighbourhood discrepancy zero.
\end{proof}

\begin{lemma}\label{clm:ND-z_i'}
For any valid subgraph $H$,
$$0 \le \ND_H(z_i',f), \ND_H(a_i',f) < 1/t^2.$$
\end{lemma}
\begin{proof}
The first inequality is immediate from the definition of neighbourhood discrepancy.  For the second inequality, observe that we only have $\ND_H(z_i',f) > 0$ if the edge $z_iz_i'$ belongs to $H$.  By validity of $H$, note that $z_i'$ must in this case have the same neighbourhood in $G$ and in $H$.  Therefore we have
\begin{align*}
\ND_H(z_i',f) = \left(4t - \frac{3t^2 \cdot 4t + t}{3t^2 + 1}\right)^2 =& \left(\frac{12t^3 + 4t - 12t^3 - t}{3t^2 + 1}\right)^2 \\ =& \left(\frac{3t}{3t^2 + 1}\right)^2 < 1/t^2.
\end{align*}
Similarly, we only have $\ND_H(a_i',f) > 0$ if the edge $a_ia_i'$ belongs to $H$.  In this case,
$$\ND_H(a_i',f) = \left(t - \frac{2t + t^3}{t^2+1}\right)^2 = \left(\frac{t^3 + t - 2t - t^3}{t^2+1}\right)^2 < 1/t^2.$$
\end{proof}

\begin{lemma}\label{clm:valid-deg-lb}
For any valid subgraph $H$,
$$\sum_{v \in V} \ln d_H(v) \ge 6n \ln t + 2n.$$
\end{lemma}
\begin{proof}
Consider first a single variable gadget, corresponding to the variable $x_i$, and note that every edge incident with a leaf must be present in $H$.  We therefore conclude that
\begin{itemize}
\item $d_H(z_i) \ge t$,
\item $d_H(z_i') \ge 3t^2$, and
\item $d_H(u_i) \ge 3t$.
\end{itemize}
It follows that the contribution to the sum from this gadget is at least 
\begin{align*}
\ln t + \ln 3t^2 + \ln 3t &\ge \ln t + 2 \ln t + \ln 3 + \ln t + \ln 3 \\ 
	& \ge 4 \ln t + 2.
\end{align*}
Now consider a single clause gadget, corresponding to the clause $C_j$.  Invoking the validity of $H$ again, we observe that $d_H(a_j') \ge t^2$, so the contribution from this gadget is at least $2\ln t$.

Summing over all variables and clauses gives the result.
\end{proof}

\begin{lemma}\label{clm:valid-deg-ub}
Let $H$ be any subgraph of $G$ (not necessarily valid).  Then
$$\sum_{v \in V} \ln d_H(v) \le 6n \ln t + 20n.$$
\end{lemma}
\begin{proof}
Note that the sum is maximised when $H$ is equal to $G$, so it suffices to show that the right-hand side gives an upper bound on $\sum_{v \in V} \ln d_G(v)$.

Note that, for $x \geq 3$, $\ln(x+1) < \ln(x+2) < \ln(x) + 1$.
For a single variable gadget the contribution to this sum is then:
\begin{align*}
\ln(3t^2 + 1) + & \ln(3t + 2) + \ln 5 + \ln (3t + 1) + 3 \ln 3 \\
	& \le 2 \ln t + \ln 3 + 1 + \ln t + \ln 3 + 1 + \ln 5 + \ln t + \ln 3 + 1 + 6 \\
	& \le 4 \ln t + 17. 
\end{align*}
For each clause gadget, the contribution to this sum is
$$\ln(t^2 + 1) + \ln 4 \le 2 \ln t + 3.$$
Summing over all variables and clauses gives
$$\sum_{v \in V} \ln d_G(v) \le  6n \ln t + 20n,$$
as required.
\end{proof}

\begin{lemma}\label{clm:sat-ND}
If $\Phi$ is satisfiable, there is a valid subgraph $H$ such that for all $v \in V \setminus \{z_i', a_i': 1 \le i \le n\}$ with $d_G(v) > 1$ we have $\ND(v,H) = 0$.
\end{lemma}
\begin{proof}
Let $b: \{x_1,\ldots,x_n\} \rightarrow \{\true,\false\}$ be a truth assignment such that, for each clause $C_j$ in $\Phi$, exactly one variable in $C_j$ evaluates to $\true$ under $b$.  We define a valid subgraph $H$ of $G$ with reference to $b$.

The subgraph $H$ contains all edges within each clause gadget (i.e. all edges incident with $a_i'$ for each $i$).  Let $G_i$ be the variable gadget corresponding to $x_i$, and $H_i$ the subgraph of $H$ induced by the same set of vertices.  If $b(x_i) = \true$, we set $H_i$ to be $G_i \setminus z_iv_i$.  If $b(x_i) = \false$, we set $H_i$ to be $G_i \setminus \{v_iw_{i,1},v_iw_{i,2},v_iw_{i,3},z_iz_i'\}$.

To complete the definition of $H$, we define the set of edges in $H$ that have one endpoint in a variable gadget and one in a clause gadget: for each $i$, $j$, and $\ell$, $H$ contains the edge $w_{i,\ell}a_j$ if and only if $b(x_i) = \true$.
It is easy to verify that $H$ is a valid subgraph; it remains to demonstrate that $\ND_H(v,f) = 0$ for the required vertices.

First consider $u_i$, for any $i$.  The neighbourhood of $u_i$ does not depend on the value of $b(x_i)$, so in all cases we have
$$\ND_H(u_i,f) = \left(7t - \frac{3t(7t+1) + 4t}{3t+1} \right)^2 = \left(\frac{21t^2 + 7t - 21t^2 - 3t -4t}{3t^2 + 1} \right)^2 = 0.$$

Now consider $v_i$, for any $i$.  If $b(x_i) = \true$, then $N_H(v_i) = \{u_i,w_{i,1},w_{i,2},w_{i,3}\}$ and
$$\ND_H(v_i,f) = \left(4t - \frac{7t + 3t + 3t + 3t}{4} \right)^2 = 0.$$
On the other hand, if $b(x_i) = \false$, then $N_H(v_i) = \{u_i,z_i\}$ and
$$\ND_H(v_i,f) = \left(4t - \frac{7t + t}{2} \right)^2 = 0.$$

Next consider $z_i$, for any $i$.  Note that $|N_H(z_i) \cap \{z_i',v_i\}| = 1$ for either value of $b(x_i)$, and moreover that $z_i$ is always adjacent to its leaf neighbours, so in both cases we have
$$\ND_H(z_i,f) = \left(t - \frac{3t(t-1) + 4t}{3t + 1} \right)^2 = \left(\frac{3t^2 + t - 3t^2 + 3t - 4t}{3t + 1} \right)^2 = 0.$$

Next consider $w_{i,j}$, for any $i$ and $j$.  In all cases, $w_{i,j}$ is adjacent to its leaf neighbour.  If $b(x_i) = \false$, this is the only neighbour of $w_{i,j}$, and it is clear that $\ND_H(w_{i,j},f) = 0$.  If, on the other hand, $b(x_i) = \true$ then $w_{i,j}$ is additionally adjacent to both $v_i$ and $a_j$ for some clause $C_j$.  In this case we have
$$\ND_H(w_{i,j},f) = \left(3t - \frac{3t + 4t + 2t}{3} \right)^2 = 0.$$

Finally, consider $a_j$, for any $j$.  By the definition of $H$, $a_j$ is adjacent to its leaf neighbours in $G$ and, since $C_j$ contains exactly one variable that evaluates to true under $b$, exactly one vertex $w_{i,\ell}$ for some values of $i$ and $\ell$; we therefore have
$$\ND_H(a_j,f) = \left(2t - \frac{t + 3t}{2}\right)^2 = 0.$$
\end{proof}

\begin{lemma}\label{clm:unsat-ND}
If $\Phi$ is not satisfiable, then for any valid subgraph $H$, there exists a vertex $v \in V \setminus \{z_i',a_i':1 \le i \le n\}$ with $d_G(v) > 1$ such that
$$\ND_H(v,f) \ge t^2/9.$$
\end{lemma}
\begin{proof}
Suppose for a contradiction that there is a valid subgraph $H$ so that for every such vertex $v$ we have $\ND_H(v,f) < t^2/9$.

We begin by arguing that $N_H(v_i) \in \{\{u_i,z_i\}, \{u_i,w_{i,1},w_{i,2},w_{i,3}\}\}$; this will allow us to construct a truth assignment based on $H$.  We first argue that we must have $u_i \in N_H(v_i)$: to see this, observe that if this is not the case then $\sum_{u \in N_H(v_i)} f(u)/d_H(v_i) \le 3t$ so we have $\ND_H(v_i,f) \ge t^2$.  Now suppose $z_i \in N_H(v_i)$.  If we also have $w_{i,j} \in N_H(v_i)$ for some $j$, then 
$$\frac{\sum_{u \in N_H(v_i)}f(u)}{d_H(v_i)} \le \frac{7t + t + 3t}{3} = \frac{11t}{3},$$ 
so $\ND_H(v_i,f) \ge (t/3)^2 = t^2/9$, a contradiction; it follows that if $z_i \in N_H(v_i)$ then $N_H(v_i) = \{u_i,z_i\}$.  Now suppose that $z_i \notin N_H(v_i)$.  In this case, if $N_H(v_i) \neq \{u_i,w_{i,1},w_{i,2},w_{i,3}\}$, we have 
$$\frac{\sum_{u \in N_H(v_i)}f(u)}{d_H(v_i)} \ge \frac{7t + 3t + 3t}{3} = \frac{13t}{3},$$
so $\ND_H(v_i) \ge t^2/9$, again giving a contradiction.

We therefore conclude that, for each $i$,  $N_H(v_i) \in \{\{u_i,z_i\}, \{u_i,w_{i,1},w_{i,2},w_{i,3}\}\}$.  We now define a truth assignment $b: \{x_1,\ldots,x_n\} \rightarrow \{\true,\false\}$ by setting
\[
b(x_i) = \begin{cases}
				\true	& \text{if $N_H(v_i) = \{u_i,w_{i,1},w_{i,2},w_{i,3}\}$,}\\
				\false	& \text{if $N_H(v_i) = \{u_i,z_i\}.$}
		 \end{cases}
\]
Since $\Phi$ is not satisfiable, there is at least one clause $C_j$ in $\Phi$ such that, under $b$, the number of variables in $C_j$ evaluating to $\true$ is either zero, two or three.

Suppose first that no variable in $C_j$ evaluates to $\true$ under $b$.  If $a_j$ is not adjacent to any vertex $w_{i,\ell}$, then $\ND_H(a_i,f) = t^2$, a contradiction.  Therefore $a_j$ is adjacent to at least one vertex $w_{i,\ell}$ where $b(x_i) = \false$.  By definition of $b$, $w_{i,\ell}$ is not adjacent to $v_i$, so we have
$$\ND_H(w_{i,\ell},f) \ge \left(3t - \frac{3t + 2t}{2}\right)^2 = t^2/4,$$
a contradiction.

We can therefore conclude that at least two variables in $C_j$ evaluate to true under $b$. Suppose two true variables appearing in $C_j$ are $x_i$ and $x_{\ell}$, and that $w_{i,r}$ and $w_{\ell,s}$ are adjacent to $a_j$ in $G$.  We begin by arguing that both $w_{i,r}$ and $w_{\ell,s}$ must be adjacent to $a_j$ in $H$.  Suppose for a contradiction (without loss of generality) that $w_{i,r}a_j \notin E(H)$.  In this case, since $b(x_i) = \true$, we know that $v_iw_{i,r} \in E(H)$, so we have
$$\ND_H(w_{i,r},f) = \left(3t - \frac{3t + 4t}{2}\right)^2 = t^2/4,$$
giving the required contradiction.  We therefore conclude that $w_{i,r}$ and $w_{\ell,s}$ are adjacent to $a_j$ in $H$, but in this case we have
$$\ND_H(a_j,f) \ge \left(2t - \frac{t + 3t + 3t}{3}\right)^2 = t^2/9,$$
again giving a contradiction and completing the proof.
\end{proof}

We now give bounds on the possible values for $\score(H,f)$ depending on whether or not $\Phi$ is satisfiable.

\begin{lemma}\label{lma:sat-bound}
If $\Phi$ is satisfiable, there is a valid subgraph $H$ with
$$\score(H,f) \ge 6 n \ln t - n \ln (12nt).$$
\end{lemma}
\begin{proof}
By Lemma \ref{clm:sat-ND}, in this case there is a valid subgraph $H$ so that $\ND_H(v,f) = 0$ for all vertices $v$ with degree greater than one in $G$, other than vertices $z_i'$ for some $i$ and $a_j'$ for some $j$.  It follows that for this choice of $H$
\begin{align*}
\sum_{v \in V}d_H(v) \ND_H(v,f) = &\sum_{v \text{ a leaf in }G} \ND(v,f) + \sum_i d_H(z_i') \ND_H(z_i',f)  \\ &+ \sum_j d_H(a_j') \ND_H(a_j',f).
\end{align*}
Applying Lemmas \ref{clm:ND-leaves} and \ref{clm:ND-z_i'}, we see that
\begin{align*}
\sum_{v \in V}d_H(v) \ND_H(v,f) & \le 6nt + n \cdot (3t^2 + 1) \cdot 1/t^2 + n \cdot (t^2 + 1) \cdot 1/t^2 \\ & \le 6nt + 4n + 2n/t^2.
\end{align*}
By Lemma \ref{clm:valid-deg-lb} and since $t\geq 1$, we have
\begin{align*}
\score(H,f) &\ge 6 n \ln t + 2n - n \ln (6nt + 4n + 2n/t^2) \\
	&> 6 n \ln t - n \ln (12nt)
\end{align*}
as required.\end{proof}

\begin{lemma}\label{lma:unsat-bound}
If $\Phi$ is not satisfiable, then for every valid subgraph $H$ we have
$$\score(H,f) \le 6n \ln t + 20n - n \ln(t^2/9).$$
\end{lemma}
\begin{proof}
This follows immediately from Lemmas \ref{clm:valid-deg-ub} and \ref{clm:unsat-ND}.
\end{proof}

We are now ready to prove Theorem~\ref{thm:hard}, which we restate here for convenience.
\theoremnphard*
\begin{proof}
We suppose for a contradiction that there is a polynomial-time algorithm $\mathcal{A}$ to solve \StatOpt on planar graphs, and show that this would allow us to solve \fancySAT\, in polynomial time.

Given an instance $\Phi$ of \fancySAT, where we will assume without loss of generality that $\Phi$ has $n > e^{47}$ variables, we proceed as follows.  First construct $(G_{\Phi},f)$ as defined above, taking $t = n^2$; it is clear that this can be done in polynomial time in $|\Phi|$.  Note that $G_{\Phi}$ is planar: to see this, observe that repeatedly deleting vertices of degree one gives a subdivision of the incidence graph which is planar by assumption.  We then run $\mathcal{A}$ on $(G_{\Phi},f)$ and return YES if the output is at least $\frac{17}{2} n \ln n$, and NO otherwise.

It remains to demonstrate that this procedure gives the correct answer.  Suppose first that $\Phi$ is satisfiable.  In this case, by Lemma \ref{lma:sat-bound}, we know that there exists a subgraph $H$ of $G$ with 
\begin{align*}
\score(H,f) &\ge 6n\ln t - n \ln(12nt)\\
		   &= 6n \ln n^2 - n \ln (12n^3) \\
		   &= 12 n \ln n - 3 n \ln n - n \ln 12 \\
		   &\ge 9 n\ln n - 3n\\
		   &> \frac{17}{2} n \ln n,
\end{align*}
since $3 < \ln n / 2$, so our procedure returns YES.

Conversely, suppose that $\Phi$ is not satisfiable.  In this case, by Lemma \ref{lma:unsat-bound} we know that, for every valid subgraph $H$ we have
\begin{align*}
\score(H,f) &\le 6n \ln t + 20n - n \ln(t^2/9)\\
		    &= 6n \ln n^2 + 20n - n \ln(n^4/9)\\
		    &= 12n \ln n + 20n - 4n \ln n + n \ln 9\\
		    &\le 8 n \ln n + 23n\\
		    &< \frac{17}{2}n \ln n,
\end{align*}
since $23 < \ln n / 2$, so our procedure returns NO.
\end{proof}

\section{Simplifications of the problem}
\label{sec:simple}

One may wonder if the hardness of \StatOpt is due to the interplay between the two parts of the objective function.
We show in Section~\ref{sec:nd-only} that just determining if there is a valid subgraph with total neighbourhood discrepancy below some given constant is NP-complete, even if the input graph is planar and has maximum degree at most five.
In Section~\ref{sec:ideal} we that show that subgraphs that have zero neighbourhood discrepancy everywhere (if they exist) can be found in time linear in the number of edges, however finding subgraphs that have zero neighbourhood discrepancy everywhere excluding leaves is NP-complete.

\subsection{Minimising neighbourhood discrepancy}\label{sec:nd-only}

Consider the following problem, which requires us to minimise only the neighbourhood discrepancy.

\begin{framed}
\noindent
\textbf{\AvNbdOpt}\\
\textit{Input:} A graph $G=(V,E)$, a function $f: V \rightarrow \mathbb{Q}$, and $k \in \mathbb{Q}$.\\
\textit{Question:} Is there a spanning subgraph $H$ of $G$ such that $d_H(v) \geq 1$ for all $v \in V$ and 
$$\sum_{v \in V} \left(f(v) - \frac{\sum_{u \in N_H(v)} f(u)}{d_H(v)}\right)^2 \quad \leq \quad k \quad ?$$
\end{framed}

First observe that the \AvNbdOpt{} is clearly in NP.
We will show the NP-hardness of \AvNbdOpt{} by giving a reduction from \fancySAT, which we used earlier in Section~\ref{sec:hardness}.
As a reminder, the \emph{incidence graph} $G_{\Phi}$ of a CNF formula $\Phi$ is a bipartite graph whose vertex sets correspond to the variables and clauses of $\Phi$ respectively, and in which a variable $x$ and clause $C$ are connected by an edge if and only if $x$ appears in $C$.
This was shown to be NP-complete in \cite{moore01}.

Let $\Phi = C_1 \wedge \cdots \wedge C_m$ be the input to an instance of \fancySAT, where each clause $C_j$ is of the form $(x_{j_1},x_{j_2},x_{j_3})$, and suppose that the variables appearing in $\Phi$ are $x_1,\ldots,x_n$.  We will construct an instance $(G,f,k)$ of \AvNbdOpt\, which is a yes-instance if and only if $\Phi$ is a yes-instance for \fancySAT.

\begin{figure}
\centering
\includegraphics[width = 0.7 \linewidth]{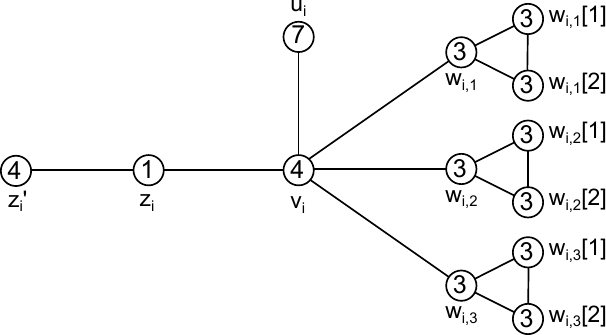}
\caption{The construction of the variable gadget $G_i$; the number in each vertex indicates the corresponding value of the function $f$.}
\label{fig:variable_gadget}
\end{figure}

The graph $G$ consists of \emph{variable gadgets} and \emph{clause gadgets}, with some edges between variable and clause gadgets.  For each variable $x_i$, we have a variable gadget $G_i$, as illustrated in Figure \ref{fig:variable_gadget}.  $G_i$ consists of 13 vertices:
\begin{itemize}
\item $v_i$, with $f(v_i) = 4$;
\item $u_i$, with $f(u_i) = 7$;
\item $z_i$, with $f(z_i) = 1$, and $z_i'$, with $f(z_i')=4$;
\item $w_{i,j}$ for $j \in \{1,2,3\}$, with $f(w_{i,j}) = 3$;
\item the \emph{triangle vertices} $w_{i,j}[\ell]$ for $j \in \{1,2,3\}$ and $\ell \in \{1,2\}$, with $f(w_{i,j}[\ell])=3$.
\end{itemize}
The gadget $G_i$ also contains the following edges:
\begin{itemize}
\item $v_iu_i$, $v_iz_i$ and $z_iz_i'$;
\item $v_iw_{i,j}$ for $j \in \{1,2,3\}$;
\item $w_{i,j}w_{i,j}[1]$, $w_{i,j}w_{i,j}[2]$ and $w_{i,j}[1]w_{i,j}[2]$ for $j \in \{1,2,3\}$.
\end{itemize}
For each clause $C_j$, we have a clause gadget $H_j$, which consists of two vertices $a_j$ and $b_j$ joined by an edge; we set $f(a_j) = 2$ and $f(b_j)=1$.

We complete the construction of $G$ by describing the edges between clause and variable gadgets.  
If the variable $v_i$ appears in clauses $C_{j_1}$, $C_{j_2}$, and $C_{j_3}$, with $j_1 < j_2 < j_3$, then we have edges $w_{i,1}a_{j_1}$, $w_{i,2}a_{j_2}$, and $w_{i,3}a_{j_3}$.
Finally, we set $k = 27n + m$.  It is clear that we can construct $(G,f,k)$ from $\Phi$ in polynomial time.

It is straightforward to verify that the maximum degree of $G$ is $5$.  To see that $G$ is planar, first note that (repeatedly) adding or removing vertices of degree one does not change the planarity of a graph. Thus $G$ is planar if and only if the graph $G'$, obtained by deleting the vertices $u_i$, $z_i'$ and $z_i$ from each variable gadget $G_i$ and $b_j$ from each clause gadget, is planar.  Moreover, it is clear that $G'$ is planar if and only if the graph $G''$, obtained from $G'$ by deleting all triangle vertices, is planar.  But $G''$ is a subdivision of the incidence graph $G_{\Phi}$ which, by assumption, is planar.  We can therefore conclude that $G$ is indeed planar.

We now argue that $(G,f,k)$ is a yes-instance if and only if $\Phi$ is a yes-instance.
Recall that $H$ is a \emph{valid} subgraph of $G$ if $V(H) = V(G)$ and $d_H(v) \geq 1$ for all $v \in V(G)$.
We now argue that the neighbourhood discrepancies of certain vertices in $V$ is independent of our choice of valid subgraphs $H$.

\begin{lemma}\label{clm:cont-variable-vxs}
Let $H$ be any valid subgraph of $G$.  Then, for all $i \in \{1,\ldots,n\}$,
\[
\ND_H(u_i,f) = \ND_H(z_i,f) = \ND_H(z_i',f) = 9.
\]
\end{lemma}
\begin{proof}
First consider $u_i$.  Since $d_G(u_i) = 1$, we know that $N_H(u_i) = \{v_i\}$, so $\ND_H(u_i,f) = (f(u_i) - f(v_i))^2 = (7-4)^2 = 9$.  Similarly, we know that $N_H(z_i') = \{z_i\}$, so $\ND_H(z_i',f) = (f(z_i') - f(z_i))^2 = (4-1)^2 = 9$.  Finally, we know that $\emptyset \neq N_H(z_i) \subseteq \{z_i',v_i\}$, where $f(z_i') = f(v_i) = 4$, so $\sum_{y \in N_H(z_i)} f(y) / d_H(z_i) = 4$.  Thus we have that $\ND_H(z_i,f) = (1 - 4)^2 = 9$.
\end{proof}

\begin{lemma}\label{clm:cont-clause-vxs}
Let $H$ be any valid subgraph of $G$.  Then, for all $j \in \{1,\ldots,m\}$, $\ND_H(b_j,f) = 1$.
\end{lemma}
\begin{proof}
Since $N_G(b_j) = \{a_j\}$, we must also have $N_H(b_j) = \{a_j\}$.  It follows that $\ND_H(b_j,f) = (f(b_j) - f(a_j))^2 = (1 - 2)^2 = 1$.
\end{proof}

\begin{lemma}\label{clm:cont-triangle-vxs}
Let $H$ be any valid subgraph of $G$.  Then, for all $i \in \{1,\ldots,m\}$, $j \in \{1,2,3\}$ and $\ell \in \{1,2\}$, $\ND_H(w_{i,j}[\ell],f) = 0$.
\end{lemma}
\begin{proof}
The claim follows immediately from the observation that, for all $y \in N_G(w_{i,j}[\ell])$ (and hence in $N_H(w_{i,j}[\ell])$), we have $f(y) = f(w_{i,j}[\ell])$.
\end{proof}

By Lemmas \ref{clm:cont-variable-vxs}, \ref{clm:cont-clause-vxs} and \ref{clm:cont-triangle-vxs}, we see that $\sum_{v\in V(H)} \ND_H(v,f) \geq 27n + m$ for any valid subgraph $H$, so $(G,f,k)$ is a yes-instance if and only if there is some valid subgraph $H$ such that 
$$\ND_H(v_i,f) = \ND_H(w_{i,1},f) = \ND_H(w_{i,2},f) = \ND_H(w_{i,3},f) = \ND_H(a_j,f) = 0,$$
for all $i \in \{1,\ldots,n\}$ and $j \in \{1,\ldots,m\}$.  We will say that such a subgraph $H$ is \emph{good}.

We first argue that, if $\Phi$ is a yes-instance, there is a good subgraph $H$.

\begin{lemma}\label{lemma:nd-only:assign-to-graph}
Suppose that there is an assignment $g: \{x_1,\ldots,x_n\} \rightarrow \{\true,\false\}$ such that every clause $C_j$ contains precisely one true variable.  Then there is a good subgraph $H$.
\end{lemma}
\begin{proof}
We construct $H$ by deleting the following edges from $G$.  In each variable gadget $G_i$ such that $g(x_i) = \true$, we delete the edges $v_iz_i$ and $w_{i,j}w_{i,j}[\ell]$ for each $j \in \{1,2,3\}$ and $\ell \in \{1,2\}$.  In each variable gadget $G_i$ such that $g(x_i) = \false$, we delete the edges $v_iw_{i,j}$ for $j \in \{1,2,3\}$ and all edges from $G_i$ to vertices in clause gadgets.  It suffices to demonstrate that $\ND_H(v,f) = 0$ for $v \in \{v_1,\ldots,v_n\} \cup \{w_{i,j}: 1 \leq i \leq n, 1 \leq j \leq 3\} \cup \{a_j: 1 \leq j \leq m\}$.

Suppose first that $v = v_i$ for some $i$.  If $g(x_i) = \true$ then 
\begin{align*}
\ND_H(v_i,f) &= \left(f(v_i) - \frac{f(u_i) + f(w_{i,1}) + f(w_{i,2}) + f(w_{i,3})}{4}\right)^2 \\
 &= \left(4 - \frac{7 + 3 + 3 + 3}{4}\right)^2 = 0, 
\end{align*}
and if $g(x_i) = \false$ then
$$\ND_H(v_i,f) = \left(f(v_i) - \frac{f(u_i) + f(z_i)}{2}\right)^2 = \left(4 - \frac{7 + 1}{2}\right)^2 = 0,$$
as required.

Now suppose that $v = w_{i,\ell}$ for some $1 \leq i \leq n$ and $1 \leq \ell \leq 3$.  If $g(x_i) = \true$ then $w_{i,j}$ has precisely two neighbours in $H$, $v_i$ and $a_j$ for some $j \in \{1,\ldots,m\}$.  Thus we have
$$\ND_H(w_{i,\ell},f) = \left(f(w_{i,\ell},H) - \frac{f(v_i) + f(a_j)}{2}\right)^2 = \left(3 - \frac{4 + 2}{2}\right)^2 = 0.$$
If, on the other hand, $g(x_i) = \false$, we have 
$$\ND_H(w_{i,\ell},f) = \left(f(w_{i,\ell}) - \frac{f(w_{i,\ell}[1]) + f(w_{i,\ell}[2]}{2}\right)^2 = \left(3 - \frac{3+3}{2}\right)^2 = 0.$$

Finally, suppose that $v = a_j$ for some $j \in \{1,\ldots,m\}$.  Since exactly one variable appearing in $C_j$ evaluates to true under $g$, there is exactly one edge from $a_j$ to a vertex belonging to a clause gadget; the unique neighbour of $a_j$ in a clause gadget will be $w_{i,\ell}$ for some $i \in \{1,\ldots,n\}$ and $\ell \in \{1,2,3\}$.  Thus
$$\ND_H(a_j,f) = \left(f(a_j) - \frac{f(b_j) + f(w_{i,\ell})}{2}\right)^2 = \left(2 - \frac{1 + 3}{2}\right)^2 = 0,$$
completing the proof that $H$ is good.
\end{proof}

Conversely, we now argue that the existence of a good subgraph $H$ implies that $\Phi$ is a yes-instance.

\begin{lemma}\label{lemma:nd-only:graph-to-assign}
Suppose that there is a good subgraph $H$.  Then there is an assignment $g: \{x_1,\ldots,x_n\} \rightarrow \{\true,\false\}$ such that every clause $C_j$ contains precisely one true variable.  
\end{lemma}
\begin{proof}
We begin by observing that, for any $i \in \{1,\ldots,n\}$, we have $\cont(v_i,H) = 0$ if and only if either $N_H(v_i) = \{u_i,z_i\}$ or $N_H(v_i) = \{u_1,w_{i,1},w_{i,2},w_{i,3}\}$.  We can therefore define an assignment $g: \{x_1,\ldots,x_n\}\rightarrow\{\true, \false\}$ by setting
\begin{equation*}
g(x_i) = \begin{cases}
				\true	& \text{if } N_H(v_i) = \{u_1,w_{i,1},w_{i,2},w_{i,3}\}\\
				\false  & \text{if } N_H(v_i) = \{u_i,z_i\}.
		 \end{cases}
\end{equation*}
We claim that, with this assignment, every clause $C_j$ must contain exactly one true variable.  To show that this is true, we suppose, for a contradiction, that the clause $C_j = (x_{j_1},x_{j_2},x_{j_3})$ does not contain exactly one true literal.  Note that, in order to have $\ND_H(a_j,f) = 0$, as $b_j \in N_H(a_j)$ for every good subgraph $H$, we must have $N_H(a_j) = \{b_j, w_{i,\ell}\}$ for some $i \in \{j_1,j_2,j_3\}$ and $\ell \in \{1,2,3\}$.

Suppose first that $g(x_{j_1}) = g(x_{j_2}) = g(x_{j_3}) = \false$.  Then, by definition of $g$, we see that $w_{i,\ell}$ is not adjacent to $v_i$ in $H$, so $f(y) \leq f(w_{i,\ell})$ for all $y \in N_H(w_{i,\ell})$; it follows that, to achieve $\ND_H(w_{i,\ell},f) = 0$, we must have $f(y) = f(w_{i,\ell})$ for all $y \in N_H(w_{i,\ell})$.  Since $f(a_j) = 2 < f(w_{i,\ell}) = 3$, it follows that $a_j$ and $w_{i,\ell}$ are not adjacent in $H$, giving the required contradiction.

Now suppose that at least two variables in $C_j$ evaluate to true; it follows that there is some variable $x_{j_{r}}$ in $C_j$ such that $g(x_{j_{r}}) = \true$ but $a_jw_{j_{r},\ell} \in E(G) \setminus E(H)$ for some $\ell \in \{1,2,3\}$.  Thus $N_H(w_{j_r,\ell}) \subseteq \{v_{j_r},w_{j_r,\ell}[1],w_{j_r,\ell}[2]\}$ and so $f(y) \geq f(w_{j_r,\ell})$ for all $y \in N_H(w_{j_r,\ell})$.  Moreover, by definition of $g$, we know that $v_i \in N_H(w_{j_r,\ell})$, where $f(v_i) > f(w_{j_r,\ell})$; this gives $\ND_H(w_{j_r,\ell},f) > 0$, a contradiction.

Thus we can conclude that every clause $C_j$ contains precisely one true variable.
\end{proof}

\begin{theorem}
\AvNbdOpt{} is NP-complete, even when restricted to input graphs $G$ that are planar and have maximum degrees at most five.
\end{theorem}
\begin{proof}
The result follows immediately from Lemmas~\ref{lemma:nd-only:assign-to-graph} and \ref{lemma:nd-only:graph-to-assign}.
\end{proof}

\subsection{Ideal and near-ideal subgraphs}\label{sec:ideal}

An obvious upper-bound to $\score(H, f)$ is given by $\sum_{v\in V(H)} \ln d_H(v)$ (i.e.~assume every vertex has zero neighbourhood discrepancy), so a natural question to ask is whether, for a given graph $G$ and function $f$, a valid subgraph $H$ of $G$ can be found  that achieves this bound.
In such a graph, it must hold that $\ND_H(v,f) = 0$ for every $v\in V(H)$.
We say such a graph $H$ is \emph{$f$-ideal} (or simply ideal, if $f$ is clear from the context).
We now show that this definition is equivalent to saying that a graph $H$ is $f$-ideal if and only the restriction of $f$ to any connected component of $H$ is a constant-valued function.

\begin{theorem}\label{lemma:ideal-constant}
A graph $H$ is $f$-ideal if and only if for each connected component $C_i$ in $H$ there exists some constant $c_i$ such that $f(v) = c_i$ for all $v\in V(C_i)$.
\end{theorem}
\begin{proof}
Let $P$ denote a path of maximal length in an $f$-ideal graph such that the weights of the vertices of $P$ strictly increase as one follows the path.
In an ideal graph, any edge between vertices of different weights means that $P$ must contain at least two distinct vertices, however the first and last vertices in such a path cannot have zero neighbourhood discrepancy.
Thus, no such path on one or more edges can exist in an ideal graph, so a graph $G$ is ideal if and only if for each connected component $C_i$ in $G$ there exists some constant $c_i$ such that $f(v) = c_i$ for all $v\in V(C_i)$.
\end{proof}

Thus, ideal subgraphs can be found by removing any edge $uv$ if $f(u) \neq f(v)$ (in $O(|E|) = O(n^2)$ time), and if necessary we can test if such a graph has no isolated vertices (and thus is valid) quickly.

The proof of Theorem~\ref{lemma:ideal-constant} relies on maximal paths with increasing weights, so one might be tempted to relax the ideal definition to only apply on vertices that are not leaves.
We therefore say a graph $H$ is {\em $f$-near-ideal} if $\ND_H(v,f) = 0$ for every $v\in V(H)$ with $d_H(v) \geq 2$.
In other words, we now allow non-zero neighbourhood discrepancy, but only at leaves.

\begin{framed}
\noindent
\textbf{\nideal} \\
\textit{Input:} A graph $G = (V,E)$ where $|V|=n$, and a function $f : V \mapsto \mathbb{Q}$.\\
\textit{Question:} Is there a valid subgraph $H$ of $G$ such that $H$ is $f$-near-ideal? 
\end{framed}
While an ideal subgraph (if one exists) can be found quickly, it turns out that solving \nideal is NP-complete, even on trees.
We reduce from subset-sum, which is NP-complete~\cite{Karp72}, and which we define as follows.
\begin{framed}
\noindent
\textbf{\ssum} \\
\textit{Input:} An integer $k$, and a set of integers $S=\{s_1,s_2,\ldots,s_n\}$.\\
\textit{Question:} Is there a subset $U\subseteq \{1,2,\ldots,n\}$ such that $\sum_{u\in U} s_u = k$? 
\end{framed}

\begin{figure}
\centering
\begin{tikzpicture}
\tikzstyle{vert} = [circle,draw, minimum size=7mm, inner sep=0pt]
\node[vert,label=above:{$v_t$}] (k) at (0,0) {$-k$};
\node[vert,label=above:{$v_s$}] (zero) at (1.5,0) {$0$};
\node[vert,label=below:{$v_z$}] (zero2) at (1.5,-1.5) {$0$};
\node[vert,label=above:{$v_1^1$}] (a1) at (3,1.5) {$s_1$};
\node[vert,label=above:{$v_1^2$}] (a2) at (4.5,1.5) {$s_1$};
\node[vert,label=above:{$v_1^3$}] (a3) at (6,1.5) {$s_1$};
\node[vert,label=above:{$v_2^1$}] (b1) at (3,0) {$s_2$};
\node[vert,label=above:{$v_2^2$}] (b2) at (4.5,0) {$s_2$};
\node[vert,label=above:{$v_2^3$}] (b3) at (6,0) {$s_2$};
\node[vert,label=below:{$v_n^1$}] (z1) at (3,-1.5) {$s_n$};
\node[vert,label=below:{$v_n^2$}] (z2) at (4.5,-1.5) {$s_n$};
\node[vert,label=below:{$v_n^3$}] (z3) at (6,-1.5) {$s_n$};

\draw (k) to (zero);
\draw (zero) to (zero2);
\draw (zero) to (a1);
\draw (a1) to (a2);
\draw (a2) to (a3);
\draw (zero) to (b1);
\draw (b1) to (b2);
\draw (b2) to (b3);
\draw (zero) to (z1);
\draw (z1) to (z2);
\draw (z2) to (z3);

\tikzstyle{dotted} = [very thick,dash pattern=on \pgflinewidth off 4pt, dash phase=2pt]
\draw[dotted] (b1) to (z1);
\draw[dotted] (b2) to (z2);
\draw[dotted] (b3) to (z3);
\end{tikzpicture}
\caption{Diagram of gadget for reduction from subset sum. The values inside the vertices are their associated weights.}\label{fig:subset-gadget}
\end{figure}
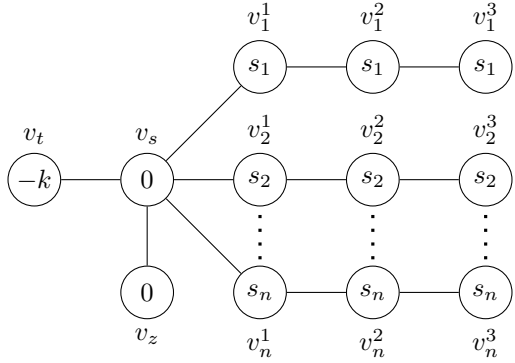

Given an instance $(S,k)$ of \ssum,  
we will construct a graph $G$ with weight function $f$ such that $(G,f)$ has a near-ideal subgraph if and only if there is a solution to our instance of subset sum. 

The graph $G$ contains $3n+3$ vertices labelled as follows:
\begin{itemize}
    \item $v_t$ for the target value, $v_s$ for a partial sum, and $v_z$ for a pendant,
    \item $v_p^j$ for $p\in \{1,\ldots,n\}$ and $j\in\{1,2,3\}$.
\end{itemize}

Vertex $v_s$ is adjacent to vertices $v_t$, $v_z$, and $v_p^1$ for $p\in\{1,\ldots,n\}$. For each $p\in\{1,\ldots,n\}$, $v_p^1$ is adjacent to $v_p^2$, and $v_p^2$ is adjacent to $v_p^3$.
This graph can be seen in Figure~\ref{fig:subset-gadget}.
We define $f$ as follows:
\begin{itemize}
    \item $f(v_t) = -k$,
    \item $f(v_s) = f(v_z) = 0$, and
    \item  $f(v^j_p) = s_p$ for $p\in\{1,\ldots,n\}$, and for $j\in\{1,2,3\}$.
\end{itemize}

Note that for the condition $d_H(v) \geq 1$ to hold for our subgraph $H$, the only edges in $G$ that might not be in $H$ are of the form $v_sv_p^1$ or $v_p^1v_p^2$ for some $p\in\{1,\ldots,n\}$.
Additionally, for any $p\in\{1,\ldots,n\}$, at most of one of $v_sv_p^1$ or $v_p^1v_p^2$ can be removed.

We are now ready to show that $G$ has a near-ideal subgraph if and only if $(S,k)$ is a yes-instance for \textsc{Subset Sum}.

\begin{theorem}
  \nideal is NP-complete, even if the input graph $G$ is a tree.
\end{theorem}
\begin{proof}

First, assume that $G$ has an $f$-near-ideal subgraph $H$.
Consider the vertex $v_s$ and its neighbours in $H$.
Since $v_s$ must have degree at least 2 in $H$, it must hold that $\ND_H(v_s,f) = 0$, giving us
\begin{align*}
    0 =& \left(f(v_s) - \frac{\sum_{w\in N_H(v_s)}f(w)}{d_H(v_s)}\right)^2.
\end{align*}
As $f(v_s)=0$ and $d_H(v_s) \geq 2$, this implies 
\begin{align*}
    0 =& \sum_{w\in N_H(v_s)}f(w).
\end{align*}
The only neighbours of $v_s$ are $v_t$, $v_z$, and for some $U\subseteq S$, vertices of the form $v_u^1$ for every $u\in U$.
This gives us
\[
    0 = f(v_t) + 0 + \sum_{u\in U}f(v^1_u),
\]
and after rearranging and substituting in values of $f$ we get
\[
    k = \sum_{u\in U} s_u,
\]
and thus $U$ is a solution to our instance of subset sum.

Now, assume that $U$ is a solution to \ssum (i.e.,~there exists a $U\subseteq S$ such that $\sum_{u\in U}s_u = k$). 
We will show that $G$ contains an $f$-near-ideal subgraph.
Let $H$ be the subgraph containing edges $v_tv_s$,
$v_sv_t$,
$v_p^2v_p^3$ for $p\in\{1,\ldots,n\}$,
$v_p^1v_p^2$ for $p\in\{1,\ldots,n\}\setminus U$,
and $v_sv_u^1$ for $u\in U$.
Note that $H$ has no isolated vertices.

The only vertices with degree at least two in $H$ are $v_s$, and vertices of the form $v_p^2$ for $p\in \{1,\ldots,n\}\setminus U$.
For any vertex of the form $v_p^2$ for $p\in \{1,\ldots,n\}\setminus U$, $\ND_H(v_p^2,f) = 0$, so the last vertex to examine is $v_s$.
As $\sum_{u\in U} s_u = k$, this implies
\begin{align*}
    \ND_H(v_s,f) =& \left(f(v_s) - \frac{\sum_{w\in N_H(v_s)}f(w)}{d_H(v_s)}\right)^2 \\
    =&  \left(0 - \frac{-k + \sum_{u\in U}f(v_z)}{d_H(v_s)}\right)^2 \\
    =&  \left(0 - \frac{-k + \sum_{u\in U}s_u}{d_H(v_s)}\right)^2 \\
    =& \left( - \frac{0}{d_H(v_s)}\right)^2 \\
    =& \; 0,
\end{align*}
and so $H$ is $f$-near-ideal.

NP-hardness of \nideal then follows from the NP-hardness of \ssum, and we obtain NP-completeness by using a near-ideal subgraph $H$ as a certificate.
\end{proof}

We note that while \ssum can be solved in pseudo-polynomial time with dynamic programming~\cite{kellerer2004subset}, this does not immediately give a pseudo-polynomial algorithm for \nideal as there is no obvious bijection between an arbitrary instance of \nideal and a graph of the form shown in Figure~\ref{fig:subset-gadget}.

\section{Parameterised Results}\label{sec:algs}
In this section we describe two parameterised algorithms for \StatOpt.
We make use of the following definitions from parameterised complexity to describe these.
A problem is in the \emph{fixed parameter tractable} (or FPT) class with respect to some parameter $k$ if the problem can be solved on inputs of size $n$ in time $f(k) \cdot n^{O(1)}$ for some computable function $f$.
Note in particular that the exponent of $n$ is constant and independent of $k$.
Another class of parameterised problems is XP: a problem is in XP with respect to some parameter $k$ if the problem can be solved on inputs of size $n$ in time $O(n^{f(k)})$.
In XP problems, the exponent of $n$ may change for different values of $k$, but if an upper bound on $k$ is given then this also upper bounds the exponent of $n$.

Both of the algorithms in this section solve the enumeration problem, rather than just the decision problem.
This means that the algorithms output a complete list of all optimal solutions to the problem.
For our problem, this will be a list of subgraphs that achieve the optimal score.
As the number of solutions may be exponential in $n$, it is not possible to bound the running time of the whole algorithm by a polynomial in $n$.
Instead, we follow Creignou et al.~\cite{Creignou2016} and use \emph{precalculation}, \emph{delay}, and \emph{postcalculation} times.
The precalculation time is the time before the first result is output,
the delay time is the time between any two successive outputs, and the postcalculation time is the time between the final output, and the termination of the algorithm.
For further background on parameterised complexity, see~\cite{cygan2015parameterized}, and for further background on parameterised enumeration, see~\cite{Creignou2016}.

Enumerating all subgraphs that optimally solve \kStatOpt allows one to investigate the uncertainty of the spatial correlation, which can then be incorporated into modelling with a Bayesian modelling framework~\cite{Lee2014}.
We define this enumeration problem as \countkStatOpt{}.

\begin{framed}
\noindent
\textbf{\countkStatOpt} \\
\textit{Input:} A graph $G = (V,E)$ where $|V| = n$, an integer $k$, and a function $f: V \rightarrow \mathbb{Q}$.\\
\textit{Output:} All spanning subgraphs $H$ of $G$, with $|E(G\setminus H)| \leq k$ and minimum degree at least one, that maximise 
$$ \score(H,f):= \sum_{v \in V}\ln d_H(v) - n \ln \left[ \sum_{v \in V} d_H(v) \left(f(v) - \frac{\sum_{u \in N_H(v)} f(u)}{d_H(v)}\right)^2\right],$$
where the maximum is taken over all spanning subgraphs $H$ of $G$ with $|E(G\setminus H)|\leq k$ and minimum degree at least one.
\end{framed}

In Section~\ref{sec:xp} we show that all solutions to \countkStatOpt 
can be enumerated with XP precalculation time (where the parameter is the maximum degree plus tree width), linear delay, and linear postcalculation time.
Then in Section~\ref{sec:low-degree} we consider the more restricted case where $G$ has maximum degree three, and show that with this restriction, all solutions to \countkStatOpt can enumerated with FPT precalculation time, linear delay and linear postcalculation time, where the parameter is the maximum number of edges that are removed.
We highlight that this restriction on the maximum degree occurs naturally in triangulations of surfaces, such as can occur when discretising spatial data~\cite{Lindgren2011}.

\subsection{An exact XP enumeration algorithm parameterised by treewidth and maximum degree}\label{sec:xp}

In this section we give an exact XP algorithm for solving \StatOpt{} on arbitrary graphs.
To aid readability, we first show that the algorithm solves the decision problem, and then at the end of this section we explain how to extend it to solve the enumeration problem \countkStatOpt.

\begin{restatable}{theorem}{theoremoptxp}\label{thm:opt-xp-maxdegree}
    \StatOpt{} can be solved in time \[
    O(2^{2\Delta(G)(tw(G)+1)}\cdot n^{2\Delta(G) + 1}).\]
\end{restatable}

The algorithm follows fairly standard dynamic programming techniques on tree decompositions, and our enumeration result follows from tracking which states lead to which.

The algorithm starts by finding a nice tree decomposition $T$ of $G$ with treewidth $tw(G)$ that is rooted at some arbitrary leaf bag.
A \emph{nice tree decomposition} is a tree decomposition with one leaf bag selected as a root bag so that the children of a bag $\nu$ are the bags adjacent to $\nu$ that are further from the root, and the additional property that each leaf bag is empty, and each non-leaf bag is either a introduce bag, forget bag, or join bag, which are defined as follows.
An introduce bag $\nu$ has exactly one child below it, say $\mu$, such that $\nu$ contains every element in $\mu$ as well as precisely one more element.
A forget bag $\nu$ has exactly one child below it, say $\mu$, such that $\nu$ contains every element in $\mu$ except one.
A join bag $\lambda$ has exactly two children below it, say $\mu$ and $\nu$, such that $\lambda$, $\mu$, and $\nu$, all have precisely the same elements.
See \cite{cygan2015parameterized}, in particular Chapter 7, for an introduction to tree decompositions, a formal definition of nice tree decompositions, and methods for constructing them.

In this section we will assume that both $n \geq 2$ and $\Delta(G) \geq 2$.
If either of these conditions does not hold, then there is at most one valid subgraph, $G$ itself, to consider so one can simply calculate and return the value of $\score(G, f)$.

We first define some specific terminology that will be useful when describing the algorithm.
Let $T$ be a tree decomposition (not necessarily nice) with an arbitrary bag labelled as the root.
For each bag $\nu \in T$, denote by $G_\nu$ the induced subgraph of $G$ consisting precisely of vertices that appear in bags below $\nu$ but do not appear in $\nu$, where we take below to mean further away from the root bag.
The set of edges between a vertex in $\nu$ and a vertex in $G_\nu$ will be important to our algorithm, so we will write $E_\nu = \{ uv \in E(G) \mid u\in \nu \wedge v \in G_\nu \}$ to be the set of edges with one endpoint in $G_\nu$ and the other in $\nu$.
An example of a graph, a tree decomposition, $G_\nu$, and $E_\nu$ are shown in Figure~\ref{fig:xp-tree-decomp}.

Our algorithm will process each bag, from the leaves towards the root, determining a set of states for each bag such that we can guarantee that the optimal solution will correspond to a state in the root bag.

%% code to draw tree decompositions, originally from https://tex.stackexchange.com/questions/356564/macro-for-rounded-polygon-around-some-nodes

%------------------------%
%---The mypoly command---%
%------------------------%

%--Getting the last Element of a list--%
\def\splicelist#1{
\StrCount{#1}{,}[\numofelem]
\ifnum\numofelem>0\relax
     \StrBehind[\numofelem]{#1}{,}[\mylast]%
\else
    \let\mylast#1%
\fi
}

%--The mypoly macro--%
%How to use:
%\myroundpoly[decorative commands]{list of names of nodes}{distance}
%list of names has to be given in clockwise order
\newcommand{\myroundpoly}[3][thin,color=black]{
%Get the last element
\splicelist{#2}
%Calculate the auxiliary coordinates for the arcs
\foreach \vertex [remember=\vertex as \succvertex
    (initially \mylast)] in {#2}{
    \coordinate (\succvertex-next) at ($(\succvertex)!#3!90:(\vertex)$);
    \coordinate (\vertex-previous) at ($(\vertex)!#3!-90:(\succvertex)$);
    \draw[#1] (\succvertex-next) --  (\vertex-previous);
}
%Draw the arcs
\foreach \vertex in {#2}{
    \tkzDrawArc[#1](\vertex,\vertex-next)(\vertex-previous)
}
}

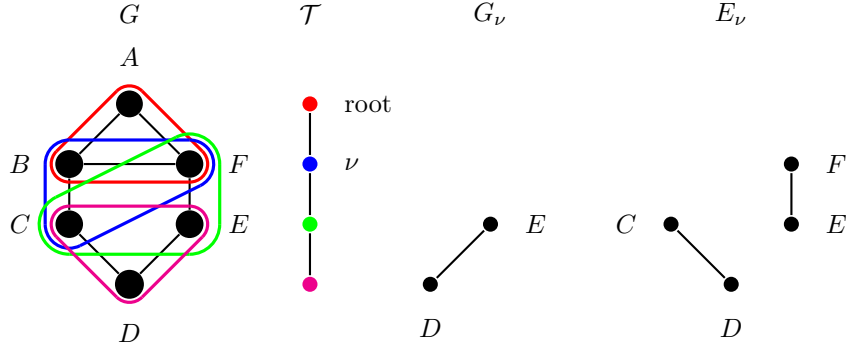
\begin{figure}
    \centering
    \begin{tikzpicture}[label distance=2mm, scale=0.8]
        \node (G) at (0,1.5) {$G$};
        \node[vert,label=above:$A$] (A) at (0, 0) {$A$};
        \node[vert,label=left:$B$] (B) at (-1, -1) {$B$};
        \node[vert,label=left:$C$] (C) at (-1, -2) {$C$};
        \node[vert,label=below:$D$] (D) at (0, -3) {$D$};
        \node[vert,label=right:$E$] (E) at (1, -2) {$E$};
        \node[vert,label=right:$F$] (F) at (1, -1) {$F$};
        \draw[edge] (A) -- (B) -- (C) -- (D) -- (E) -- (F) -- (A);
        \draw[edge] (B) -- (F);
        \myroundpoly[color=red,very thick] {B,A,F}{3mm}
        \myroundpoly[color=blue,very thick] {C,B,F}{4mm}
        \myroundpoly[color=green,very thick] {E,C,F}{5mm}
        \myroundpoly[color=magenta,very thick] {D,C,E}{3mm}
        
        \node (T) at (3,1.5) {$\mathcal{T}$};
        \node[vert,red,label=right:root] (R) at (3,0) {};
        \node[vert,blue,label=right:$\nu$] (BL) at (3,-1) {};
        \node[vert,green] (G) at (3,-2) {};
        \node[vert,magenta] (M) at (3,-3) {};
        \draw[edge] (R) -- (BL) -- (G) -- (M);
        
        \node (Gnu) at (6,1.5) {$G_\nu$};
        \node[vert,label=right:$E$] (EE) at (6,-2) {};
        \node[vert,label=below:$D$] (DD) at (5, -3) {};
        \draw[edge] (EE) -- (DD);
        
        \node (Enu) at (10,1.5) {$E_\nu$};
        \node[vert,label=right:$F$] (FFF) at (11,-1) {};
        \node[vert,label=left:$C$] (CCC) at (9,-2) {};
        \node[vert,label=right:$E$] (EEE) at (11,-2) {};
        \node[vert,label=below:$D$] (DDD) at (10, -3) {};
        \draw[edge] (DDD) -- (CCC);
        \draw[edge] (FFF) -- (EEE);
        
    \end{tikzpicture}
    \caption{From left to right, we have a graph $G$ with a tree decomposition displayed by circling vertices, the tree indexing a tree decomposition of $G$ drawn as a graph with the root and the bag $\nu$ labelled, the graph $G_\nu$ consisting of the induced subgraph on vertices $D$ and $E$, and the the set of edges $E_\nu = \{CD, EF\}$ (i.e., the edges of $G$ that are between a vertex in $G_\nu$ and a vertex in $\nu$). }
    \label{fig:xp-tree-decomp}
\end{figure}

Given a bag $\nu$ of a tree decomposition and a set of edges $I \subseteq E_\nu$, define $\mathcal{G}'_{\nu,I}$ to be the set of graphs $G'$ with:
\begin{itemize}
    \item $V(G') = V(G_\nu) \cup \nu$,
    \item $E(G') \subseteq E(G)$, and
    \item for any edge $uv \in E_\nu$, $uv \in E(G')$ if and only if $uv \in I$.
\end{itemize}

\begin{definition}\label{dfn:xp:states}
For a bag $\nu$, the set of all valid states at $\nu$ is 
    \begin{align*}
    S_\nu = \{ (I, D) \mid & I \subseteq E_\nu \textrm{, } 
    \mathcal{G}'_{\nu, I} \neq \emptyset \textrm{, and there exists a graph } H \in \mathcal{G}'_{\nu,I} \\
    & \textrm{ with } D = \sum_{v\in G_\nu} \ln d_H(v),
    \textrm{ and } d_H (v) \geq 1 \;\forall v \in V(G_\nu)\}.\end{align*}
\end{definition}

We often discuss the graph $H$ that corresponds to a state, which we now define.

\begin{definition}\label{dfn:xp:correspond}
A graph $H$ corresponds to a state $(I,D)$ if $H\in \mathcal{G}'_{\nu,I}$, $D = \sum_{v\in G_\nu} \ln d_H (v)$, and $d_H(v) \geq 1 \;\forall v\in V(G_\nu)$.
\end{definition}
Note that a graph may correspond to at most one state, while there may be multiple graphs which all correspond to the same state.

In this section 
we use $G$ (respectively $G_\nu$) to refer to the original graph in the problem (respectively an induced subgraph thereof, as defined earlier), and we will use $H$, or $H_\nu$ for a given bag $\nu$, to refer to a graph that is a subgraph of $G_\nu$ corresponding to a state.

Observe that if we are given a graph $H$, and the subgraph of $H$ induced by $V(G_\nu)$ is $H_\nu$, then for any bag $\nu$ the only information on vertices in $H_\nu$ that we need to calculate $\score (H, f)$ is $I$, $D$, and the value of $\sum d_{H_\nu}(v) \ND_{H_\nu}(v, f)$ over all vertices $v$ in $H_\nu$.
Thus for each state $(I, D)$ we will store the minimum value of $\sum_{v\in H_\nu} d_{H_\nu}(v) \ND_H(v,f)$ over subgraphs $H$ corresponding to $(I, D)$.
In our algorithm, we will represent this as a function $N: S \rightarrow \mathbb{R}$.

    We now upper bound the number of states in any one bag.
    
    \begin{lemma}\label{lemma:dp:bagsize} 
    Each bag will have at most $2^{\Delta(G)(tw(G)+1)} \cdot n^{\Delta(G)}$ states.
    \end{lemma}
    \begin{proof}
    We see that $I$ may contain up to $\Delta(G)$ edges per vertex $v \in \nu$, giving at most $2^{\Delta(G)(tw(G)+1)}$ possibilities for $I$.
    Additionally, each vertex of any subgraph $H_\nu$ corresponding to a given state (there may be at most $n$ such vertices) has a degree of at most $\Delta(G)$, giving $n^{\Delta(G)}$ possibilities for $D$ and thus the result.
    \end{proof}
    
    We now discuss the process of determining states at each type of bag in a nice tree decomposition (leaf, introduce, forget, and join).
    We begin by observing that there is only one state $(\emptyset, 0)$ at each leaf bag (apart from the root), and that the states at an introduce bag $\nu$ are exactly the states of the child of $\nu$.
    This leaves forget and join bags, which are slightly more involved.
    
We now give the algorithm for determining states at a forget bag.
For a bag that forgets vertex $v$, 
this involves taking each state at the child of $v$ and each subset of edges incident to $v$.
If the state (which contains information about edges between vertices already forgotten, and vertices not yet forgotten) is consistent with a particular subset of edges (i.e., there is no edge incident to $v$ such that the state assumes the edge will be removed, but the edge is present in the subset) then a new state is created at the forget bag.

\begin{algorithm}[htb]
\caption{Process forget bags}\label{alg:dp:forget}
\KwData{A forget bag $\nu$ that forgets vertex $v$, and the states $S_\mu$ of its child $\mu$}
\KwResult{The set $S_\nu$ of states in $\nu$}
Let $S_\nu = \emptyset$\;
Let $B$ be the set of edges in $G$ incident with $v$\;
\ForEach{state $(I_\mu, D_\mu) \in S_\mu$\label{line:xp:forget:states_below}}{ 
  \ForEach{$E\subseteq B$ with $|E| \geq 1$\label{line:xp:forget:edges}}{
      \If{$E$ is consistent with $I_\mu$ (i.e.,~for each $u\in \nu$, $uv\in E$ if and only if $uv\in I_\mu$\label{line:xp:forget:condition}}{
        Let $I_\nu = (I_\mu \cup E) \setminus \{uv \mid u \in G_\nu \}$ and $D_\nu = D_\mu + \ln |E|$\;
        Let $N(I_\nu,D_\nu) = N(I_\mu, D_\mu) + |E| \cdot \ND_{G\setminus E}(v,f)$\;
          \If(\tcp*[f]{This state has already been found}){$(I_\nu, D_\nu)\in S_\nu$}{
            Let $N(I_\nu, D_\nu) = \min \{ N(I_\nu, D_\nu), N_\nu \}$\;
          }
          \Else(\tcp*[f]{This is a new state}){
            Add $(I_\nu, D_\nu)$ to $S_\nu$\;\label{line:xp:forget:add}
            Let $N(I_\nu, D_\nu) = N_\nu$\;
          }
      }
  }
}
\Return{$S_\nu$}\;
\end{algorithm}

\begin{lemma}\label{lemma:dp:forget}
    Running Algorithm~\ref{alg:dp:forget} on a forget bag $\nu$ correctly determines the possible states $S_\nu$ as per Definition~\ref{dfn:xp:states}.
\end{lemma}
\begin{proof}
We will first show that any state in $S_\nu$ (as per Definition~\ref{dfn:xp:states}) will be found by Algorithm~\ref{alg:dp:forget}, and then show that any state found by the algorithm is valid per Definition~\ref{dfn:xp:states}.
Following the variable labelling of Algorithm~\ref{alg:dp:forget}, let $v$ be the vertex forgotten and let $\mu$ be the child of $\nu$.

Let $s_\nu=(I_\nu,D_\nu) \in S_\nu$ be a state at $\nu$ as per Definition~\ref{dfn:xp:states}.
We will show that Algorithm~\ref{alg:dp:forget} will find $s_\nu$.
Let $H$ be a subgraph corresponding to $s_\nu$, and let $H_\mu$ be the induced subgraph of $H$ with $V(H_\mu) = V(H) \setminus \{v\}$.
Letting $I_\mu = E(H_\mu) \cap E_\mu$, we see that $H_\mu \in \mathcal{G}'_{\mu,I_\mu}$, so  there must be some state $s_\mu = (I_\mu, D_\mu)$ in $S_\mu$ corresponding to $H_\mu$, and as Line~\ref{line:xp:forget:states_below} iterates over all states in $S_\mu$, the state $s_\mu$ must be considered by Algorithm~\ref{alg:dp:forget}.
Let $E$ be the set of edges in $H$ incident with $v$.
By the definition of $H$ corresponding to $s_\nu$, it must be that $|E| \geq 1$.
It follows that Line~\ref{line:xp:forget:edges} must consider the subset $E$ as a set of edges, and as both $E$ and $I_\mu$ were defined from $E(H)$, this set must satisfy the condition on Line~\ref{line:xp:forget:condition}.
Therefore it must be that Algorithm~\ref{alg:dp:forget} calculates and adds $(I_\nu, D_\nu)$ to $S_\nu$, and calculates and updates $N(I_\nu, D_\nu)$.

To see that no extraneous states are found, we consider an arbitrary state $(I_\nu, D_\nu)$ generated by Algorithm~\ref{alg:dp:forget}.
It must be added at Line~\ref{line:xp:forget:add}, so there is some $(I_\mu, D_\mu)\in S_\mu$ and some $E\subseteq E_\nu$ 
such that $uv \in E$ if and only if $uv\in I_\mu$ for each $u\in\nu$.
Then there must be some graph $H_\mu \in \mathcal{G}'_{\mu, I_\mu}$ that corresponds to $(I_\mu, D_\mu)$.
Now consider the graph $H_\nu = H_\mu \cup E$.
Any edge $uw\in E(H_\nu)$ with $u\in G_\nu$, $w\not\in G_\nu$ and $w\neq v$ must be in $I_\mu$ as $(I_\mu, D_\mu) \in S_\mu$, and such edges are then in $I_\nu$.
Any edge $uv\in E(H_\nu)$ with $u\in G_\nu$ is clearly not in $I_\nu$, and any edge $vw\in E(H_\nu)$ with $w\not\in G_\nu$ is in $I_\nu$.
As no other edges are added to $I_\nu$, we see that $I_\nu \subseteq E_\nu$, so $H_\nu \in \mathcal{G}'_{\nu, I_\nu}$.
We now only have to check that $\deg u \geq 1$ for all $u\in V(H_\nu)$.
For any $u\in V(H_\mu)$, this holds as $(I_\mu, D_\mu)\in S_\mu$, and as $V(H_\nu) \setminus V(H_\mu) = \{v\}$, and $|E| \geq 1$ by Line~\ref{line:xp:forget:edges}, we see that $(I_\nu, D_\nu)$ is a state in $S_\nu$ by Definition~\ref{dfn:xp:states}.
\end{proof}

\begin{algorithm}[htb]
\caption{Handling join bags}\label{alg:dp:join}
\KwData{A join bag $\lambda$ and the sets of states $S_\nu, S_\mu$ of its two children $\nu$ and $\mu$}
\KwResult{The set $S_\lambda$ of states in $\lambda$}
Let $S_\lambda = \emptyset$\;
\ForEach{state $(I_\nu, D_\nu) \in S_\nu$\label{line:xp:join:nu}}{ 
    \ForEach{state $(I_\mu, D_\mu) \in S_\mu\label{line:xp:join:mu}$}{
        Let $I_\lambda = I_\nu \cup I_\mu$, $D_\lambda = D_\nu + D_\mu$, and $N_\lambda = N_\nu + N_\mu$ \;\label{line:xp:join:il}
          \If(\tcp*[f]{This state has already been found}){$(I_\lambda, D_\lambda)\in S_\lambda$}{
            Let $N(I_\lambda, D_\lambda) = \min \{ N(I_\lambda, D_\lambda), N_\lambda \}$\;
          }
          \Else(\tcp*[f]{This is a new state}){
            Add $(I_\lambda, D_\lambda)$ to $S_\lambda$\;\label{line:xp:join:add}
            Let $N(I_\lambda, D_\lambda) = N_\lambda$\;
          }
    }
}
\Return{$S_\lambda$}\;
\end{algorithm}

 \begin{lemma}\label{lemma:dp:join}
    Running Algorithm~\ref{alg:dp:join} on a join bag $\lambda$ correctly determines the possible states of $\lambda$ as per Definition~\ref{dfn:xp:states}.
\end{lemma}

\begin{proof}
We again proceed by first showing that Algorithm~\ref{alg:dp:join} does correctly find every valid state, and then show that any state found by Algorithm~\ref{alg:dp:join} is valid.
Let $\nu$ and $\mu$ be the two children of $\lambda$.

Consider a state $s_\lambda=(I_\lambda,D_\lambda)\in S_\lambda$.
We want to show that Algorithm~\ref{alg:dp:join} does indeed find $s_\lambda$.
Let $H_\lambda$ be a subgraph of $G_\lambda$ corresponding to $s_\lambda$,
let $H_\nu$ be the induced subgraph of $H$ with $V(H_\nu) = V(G_\nu)$, and
let $H_\mu$ be the induced subgraph of $H$ with $V(H_\mu) = V(G_\mu)$.

For any edge $uv \in I_\lambda$ (where $v\in G_\lambda$  and $u\in \lambda$), it must be that $v\not\in \lambda$.
Then $v$ must be forgotten by some bag below $\lambda$, and as each vertex is forgotten exactly once in a nice tree decomposition, without loss of generality we can say that $v$ must have been forgotten in the subtree rooted at $\nu$.
From the definition of a nice tree decomposition, this means that $v\not\in \nu$ as well while $u\in\nu$, so $uv \in I_\nu$.
By considering all edges in $I_\lambda$ in turn, we see that $I_\lambda$ can be partitioned into $I_\nu$ and $I_\mu$, where $I_\nu$ only contains edges of the form $uv$ where $v$ was forgotten in the subtree rooted at $\nu$ and $I_\mu$ only contains edges of the form $uw$ where $w$ was forgotten in the subtree rooted at $\mu$.
As Algorithm~\ref{alg:dp:join} considers each possible state in $S_\nu$ (Line~\ref{line:xp:join:nu}) and each possible state in $S_\mu$ (Line~\ref{line:xp:join:mu}), Algorithm~\ref{alg:dp:join} will find $I_\lambda$.
Let $D_\nu = \sum_{v\in V(H_\nu)} \ln \deg_{H_\nu} v$, and let $D_\mu = \sum_{v\in V(H_\mu)} \ln \deg_{H_\mu} v$.
Then $(I_\nu, D_\nu)$ is a state of $\nu$, and
$(I_\mu, D_\mu)$ is a state of $\mu$.
As $V(G_\lambda) = V(G_\nu) \cup V(G_\mu)$ and $V(G_\nu) \cap V(G_\mu) = \emptyset$, we have $D_\lambda = D_\nu + D_\mu$ and so Algorithm~\ref{alg:dp:join} does create and store $(I_\lambda, D_\lambda)$.
Additionally, as each vertex forgotten below $\lambda$ is either forgotten at a bag below $\mu$ or forgotten at a bag below $\nu$ (but not both), we see that the value of $N(I_\lambda, D_\lambda)$ is correctly calculated and stored.

We now show that any state found by Algorithm~\ref{alg:dp:join} is valid according to Definition~\ref{dfn:xp:states}.
Let $s_\lambda = (I_\lambda, D_\lambda)$ be an arbitrary state created by Algorithm~\ref{alg:dp:join}.
It must be added at Line~\ref{line:xp:join:add}, so the algorithm will consider two valid states $(I_\nu, D_\nu)$ and $(I_\mu, D_\mu)$ of $S_\nu$ and $S_\mu$ respectively, where $I_\lambda = I_\nu \cup I_\mu$ 
and $D_\lambda = D_\nu + D_\mu$.
Consider the graphs $H_\nu$ and $H_\mu$ corresponding to to $(I_\nu, D_\nu)$ and $(I_\mu, D_\mu)$ respectively.
As we have a tree decomposition, we know that these two graphs are vertex disjoint.
Any edge $uv \in I_\lambda$ must therefore without loss of generality satisfy $u\in G_\nu$ and $v\not \in G_\nu$.
However, as $\lambda$ is a join bag, then $v\not\in G_\lambda$, and so we have $I_\lambda \subseteq E_\lambda$.
Let $H_\lambda$ be the union of $H_\nu$ and $H_\mu$, then as $H_\nu$ and $H_\mu$ are vertex-disjoint and $V(H_\lambda) = V(H_\nu) \cup V(H_\mu)$, we must have $\deg_{H_\lambda} v \geq 1$ for any $v\in V(H_\lambda)$.
We also see that $H_\lambda \in \mathcal{G}'_{\lambda, I_\lambda}$, and as $H_\nu$ and $H_\mu$ are vertex-disjoint we also know that $D_\lambda = D_\nu + D_\mu = \sum_{v\in V(H_\lambda)} \ln \deg v$, so
$s_\lambda$ corresponds to $H_\lambda$ and $s_\lambda$ is a valid state as per Definition~\ref{dfn:xp:states}.
\end{proof}

\begin{algorithm}[htb]
\caption{Find the score of an optimal subgraph}\label{alg:dp:complete}
\KwData{A graph $G$ with maximum degree $\Delta$, a nice tree decomposition $T$ of $G$ with width $tw$, and a function $f: V(G) \mapsto \mathbb\{Q\}$}
\KwResult{An optimal solution to \StatOpt{}}
Let $S_\lambda = \emptyset$\;
\ForEach{bag $\nu \in T$}{ 
    Calculate the states in $\nu$, depending on the type of $\nu$.
}
\ForEach{state $(I, D)$ in the root bag}{
    Let $N'(I, D) = D - n \ln N(I,D)$\;
}
\Return{The maximum value of $N'(I, D)$ over all states in the root bag}\;
\end{algorithm}

We are now ready to prove Theorem~\ref{thm:opt-xp-maxdegree}, which we restate here for convenience.
\theoremoptxp*
\begin{proof}
We see that Algorithm~\ref{alg:dp:complete} will calculate an optimal solution to \StatOpt{}.
The correctness of this algorithm follows from Definition~\ref{dfn:xp:states}, Lemma~\ref{lemma:dp:forget}, and Lemma~\ref{lemma:dp:join},
so we now determine the complexity of Algorithm~\ref{alg:dp:complete}.
By Lemma~\ref{lemma:dp:bagsize}, the number of states at any bag is $2^{O(\Delta(G) tw(G))} \cdot O(n^{\Delta(G)})$.
We also easily see that it is trivial to calculate the states of a leaf bag, or an introduce bag.
Algorithm~\ref{alg:dp:forget} at bag $\nu$, when forgetting $v$, considers each possible neighbourhood of $v$ (there are $O(2^{\Delta(G)})$ such neighbourhoods) in combination with each state of the child of $\nu$, so Algorithm~\ref{alg:dp:forget} runs in $O(2^{\Delta(G)}\cdot 2^{\Delta(G)(tw(G)+1)} \cdot n^{\Delta(G)}) = O(2^{\Delta(G)(tw(G)+2)} \cdot n^{\Delta(G)})$ time.

Algorithm~\ref{alg:dp:join} considers all possible pairs of states from each child, giving a runtime of $O(2^{2\Delta(G)(tw(G)+1))}\cdot n^{2\Delta(G)})$.
As a nice tree decomposition on a graph with $n$ vertices has $O(n)$ bags~\cite{cygan2015parameterized}, Algorithm~\ref{alg:dp:complete} therefore runs in $O(2^{2\Delta(G)(tw(G)+1)} \cdot n^{2\Delta(G)+1})$ time.
\end{proof}

To be able to enumerate all solutions, we need only track which states lead to which while processing bags.
This adds only a constant to the running time of our algorithms, and
creates, for any state $(I,D)$ corresponding to an optimal solution, a tree rooted at $(I,D)$ such that the children of a state $s$ correspond precisely to those states that can lead to $s$.
All optimal solutions can then be found and output by taking each such tree corresponding to an optimal solution, and enumerating all paths from the root to the leaves.
The height of this tree is at most $tw(G)$, but the reconstruction of each optimal subgraph takes $O(n^2)$ as $G$ (and indeed $H$) may have $O(n^2)$ edges.

\begin{theorem}\label{thm:tree-width-max-degree-enum}
   For an integer $k\geq 1$, all solutions to \countkStatOpt can be output on graphs with a precalculation time of $O(2^{2\Delta(G)(tw(G)+1)} \cdot n^{2\Delta(G)+1})$, with $O(n^2)$ delay, and with $O(n^2)$ postcalculation time.
\end{theorem}

\subsection{Parameterisation by edges removed in graphs with maximum degree three}\label{sec:low-degree}

We now study the problem when $G$ has maximum degree three and we want to bound the maximum number of edges that can be removed.
Restricting $G$ to graphs of maximum degree three is of interest as the dual graph of any triangulation has maximum degree three and triangulations can be used to represent discretised surfaces~\cite{he2018assessment,Lindgren2011,mindell2017using}.

\begin{restatable}{theorem}{thmfptkdeletionenum}\label{thm:fpt-k-deletion-enum}
   For an integer $k\geq 1$, all solutions to \countkStatOpt can be output on graphs with maximum degree three with a precalculation time of $2^{k(2 \log k + O(1))} n \log n$, with $O(n)$ delay, and with $O(n)$ postcalculation time.
\end{restatable}

We achieve this result by considering in turn each possibility for the graph consisting of deleted edges, and for each such graph we consider in turn the possibilities of the degree sequence of the remaining graph.
The number of distinct graphs $R$ that must be considered, and the number of degree sequences of $G\setminus R$, both depend only on $k$ and are independent of $n$.
As $R$ has maximum degree two and therefore consists only of paths and cycles, it has treewidth at most two.
We can therefore adapt well-known colour-coding methods~(see \cite[Section 13.3]{Flum2006book} for more details) for finding subgraphs with bounded treewidth in FPT time so that we can identify a subgraph $R$ in $G$ whose removal gives the biggest improvement to the neighbourhood discrepancy term while still maintaining the correct degree sequence of $G\setminus R$.
By storing the appropriate data when finding these optimal subgraphs with bounded treewidth, and also tracking which subgraphs $R$ and which degree sequences of $R$ lead to optimal values of the score function, we are able to rebuild all embeddings that achieve this optimum.

We begin with some notation.
As $R$ has maximum degree 2, 
we will let $R = \bigcup_{a=1}^t R^a$ where each $R^a$ is connected (i.e., each $R^a$ is either a path or a cycle).
For each $a$, let the vertices of $R^a$ be $\{r^a_1,\ldots,r^a_\ell\}$ such that for $j\in\{1,\ldots,m-1\}$ there is an edge from $r^a_j$ to $r^a_{j+1}$ (if $R^a$ is a cycle, there is also an edge from $r^a_1$ to $r^a_\ell$).
For an integer $j$, let $R^a_j$ be the induced subgraph of $R^a$ on vertices $\{r^a_1,\ldots,r^a_j\}$.
We will use $\psi$ to denote proper embeddings of $R^a$ (or subgraphs thereof) into $G$ (i.e.,~it must hold that for any $u,v \in V(R)$, $uv\in E(R) \implies \psi(u)\psi(v) \in E(G)$).
We will abuse notation slightly to say, for an arbitrary subgraph $R'$ of $R$, that $E(\psi(R')) = \{ \psi(u)\psi(v) \mid uv \in E(R')\} $ is the set of edges of $G$ onto which the edges of $R'$ are mapped under $\psi$.

Recalling that the neighbourhood discrepancy portion of the objective function has a negative coefficient (i.e.,~is a penalty), for a set of edges $F \subseteq E(G)$ let
\[
I(v, F) := D_G(v) \ND_G (v,f) -  d_{G\setminus F}(v) \ND_{G\setminus F} (v,f).
\]
In other words, $I(v, F)$ is the improvement to the neighbourhood discrepancy portion of our objective function from the vertex $v$ that is obtained by removing from $G$ the edges that are in the set $F \subset E(G)$.
Note that if $F$ contains edges not incident to $v$, then those additional edges do not alter the value of $I(v,F)$.

Our first result shows that the number of subgraphs $R$ that must be considered is at worst exponential in $k$ and independent of $n$.
This bound is not tight, but sufficient for our purpose.
\begin{lemma}\label{lemma:deletion:subgraphs}
  Given an integer $k$, there are $O(2^k e^{\sqrt{k}})$ non-isomorphic graphs $R$ such that $|E(R)| \leq k$ and $\Delta(R) \leq 2$.
\end{lemma}
\begin{proof}
The number of integer partitions of $k$, and therefore an upper bound on the number of ways to group $k$ edges into components, is $O(e^{\sqrt{k}})$ by~\cite{andrews1976theory}.
Any such grouping has at most $k$ components, and each component must be either a cycle or a path, giving  $O(2^k e^{\sqrt{k}})$ as required.
\end{proof}

We note that integer partitions of $k$, and thus non-isomorphic graphs $R$, can be generated in the obvious way in time exponential in $k$, which is sufficient for our purposes.

We now bound the number of degree sequences that $H$ can have, given a graph $R$ on at most $k$ vertices, if $H = G \setminus E(\psi(R))$ for some embedding $\psi$ of $R$, and such that $H$ has minimum degree at least one.
Once we have such a degree sequence, we know exactly what the first term of the objective function will be, and can focus on minimising the penalty.

\begin{lemma}\label{lemma:deletion:degreesequences}
Let $G$ be a graph with $\Delta(G) = 3$ and no isolated vertices, and let $k$ be a positive integer. Let $R$ be a graph with no isolated vertices that satisfies $|E(R)| \leq k$ and $\Delta(R) \leq 2$.
Then, there are $O(k)$ possible degree sequences for $G\setminus E(\psi(R))$ such that $G\setminus E(\psi(R))$ has no isolated vertices, where $\psi$ is an embedding of $R$ into $G$.
\end{lemma}
\begin{proof}
Clearly $G\setminus E(\psi(R))$ can only have vertices of degree one, two, or three, so for our given graphs $G$ and $R$, we must determine how many vertices of degree one, two, or three, can be in $G\setminus E(\psi(R))$.
To avoid isolated vertices, each degree two vertex of $R$ must be mapped to a degree three vertex of $G$.
What remains in $R$ is $O(k)$ vertices of degree one, and these may be mapped either to a vertex of degree two in $G$, or a vertex of degree three in $G$.
Thus the only choice is the number of vertices of degree one in $R$ that are mapped to vertices of degree three in $G$, and there are at most $O(k)$ such choices, giving our result.
\end{proof}

We note that these degree sequences can be enumerated in time exponential in $k$ in the obvious manner, and that this is sufficient for our algorithm.

We now introduce colour-coding as a technique for finding substructures within graphs; further background on colour-coding can be found in~\cite[Chapter 13]{Flum2006book}.
This involves assigning colours to both $R$ and $G$, and then searching for embeddings of $R$ into $G$ such that the embedding preserves colours.
Our result, as is common in colour-coding, relies on the existence of a suitable family of colourings; we obtain the required results from~\cite{NaorDerandomise}.
In particular, we colour each vertex in $R$ with a unique colour, and we use the same set of $|V(R)|$ colours to colour the vertices of $G$.
Note that the colouring of $G$ need not be proper.
We will say an embedding $\psi$ of $R$ into $G$ is \emph{colour-preserving} if, for reach $v\in V(R)$, the colours of $v$ and $\psi(v)$ are the same.

Recall that $R^a$ denotes a component of $R$, and that $R^a$ is on vertices $\{r^a_1, \ldots, r^a_\ell\}$ such that there is an edge from $r^a_j$ to $r^a_{j+1}$ for $j\in\{1,\ldots,m-1\}$, as well as the edge $r^a_1r^a_\ell$ if $R^a$ is a cycle.
We search for each path and cycle independently: the colouring of $R$ ensures that a set of colour-preserving embeddings of individual paths and cycles can be combined in the obvious way to give an injective (colour-preserving) mapping from $R$ into $G$.

To achieve this, we now define the problem of finding all embeddings of a vertex-coloured cycle into a vertex-coloured graph such that the improvement to the neighbourhood discrepancy is maximised.
We later explain how to adjust it to find paths as well.
Note that this process cannot necessarily determine how all vertices of degree one in $R$ are mapped onto vertices in $G$ to obtain our desired degree sequence of $G \setminus E(\psi(R))$; this detail is taken care of in our final result.

\begin{framed}
\noindent
\textbf{\enumEmbedCycle} \\
\textit{Input:} A vertex-coloured cycle $R^a$ on $|V(R^a)|$ distinct colours, and a (not necessarily proper) vertex-coloured graph $G$ on $n$ vertices, coloured with the same set of $|V(R^a)|$ colours, and with $\Delta(G) \leq 3$.\\
\textit{Output:} The colour-preserving embeddings that maximise the value of 
\[ \sum_{j=1}^{a}I(\psi(r^a_j), E(\psi(R^a)),\]
taken over all embeddings $\psi$ of $R^a$ into $G$.
\end{framed}

We solve \enumEmbedCycle in a 2-stage process. First, Algorithm~\ref{alg:embed-cycle} finds the optimal value of the improvement to the score using dynamic programming, and while doing so stores the \emph{states} (defined below) which lead to these optimal embeddings.
Then a simple backtracking algorithm can process these states and output the embeddings that achieve this optimal score.
We keep the two algorithms separate, as when solving \countkStatOpt we check numerous subgraphs $R$ and for each $R$ numerous colourings of $G$. However, not all of these subgraphs $R$ and colourings of $G$ may lead to optimal score values, so in the cases where these choices do not lead to optimal solutions, enumerating embeddings is not necessary, and may indeed cost too much in terms of time complexity to achieve our desired goal.

For $i\geq 2$, at each vertex $v$ of $G$ with colour $c_i$, Algorithm~\ref{alg:embed-cycle} stores states $(u,v,x,y)$ such that there is an embedding $\psi$ of $R^a_j$ with $\psi(r^a_1) = x$, $\psi(r^a_{i-1}) = u$, $\psi(r^a_i) = v$, and $\psi(r^a_\ell) = y$.
We say that such an embedding $\psi$ corresponds to the given state.
In a state, the edge $uv$ represents the ``previous edge'' used to reach the vertex $v$, as calculating any change to the neighbourhood discrepancy at $v$ requires us to know which earlier edges have been removed.
Similarly, to calculate the improvement at $x$, to which $r^a_1$ is mapped, we need to know how the edge $r^a_1r^a_\ell$ is mapped, and this depends on the choice of $\psi(r^a_\ell)=y$.
This will then affect the improvement at vertex $y$, so it must be tracked.

Additionally, for each vertex $v$ and each state $(u,v,x,y)$ we also store $P_v(u,v,x,y)$, the set of preceding states that lead to this state, as well as
$I_v'(u,v,x,y)$, the best value of the sum of the improvement to the neighbourhood discrepancies over the first $i-1$ vertices of the embedding of $R^a$.
Letting $\Psi$ be the set of all embeddings corresponding to $(u,v,x,y)$, the best improvement can be expressed as 
\[
I_v'(u,v,x,y) = \max_{\psi \in\Psi} \sum_{j=1}^{i-1} I(\psi(r^a_j), E(\psi(R^a))).\label{eq:imp_fun}
\]

Algorithm~\ref{alg:embed-cycle} calculates $I_v'(s)$ and $P_v(s)$ for each vertex $v$ and each relevant state $s$ at $v$ using common dynamic programming approaches.
Then the final step looks at all vertices $v$ of colour $c_\ell$ (i.e., those vertices that are picked last by Algorithm~\ref{alg:embed-cycle}) and calculates as $I_b$ the best improvement over all states $s\in S(v)$ over all such vertices $v$, after adding the improvement at vertex $v$, and also stores as $P_b$ all states that attain $I_b$.

Note that no states are stored at any vertex coloured $c_1$, and the improvement stored at a state of $v$ does not include the actual improvement of the vertex $v$, as this depends on how further edges of $R^a$ are embedded.

Keeping track of the previous states that led to the optimal value at each state allows us to backtrack to enumerate all embeddings that achieve the optimal overall value.

We will see later that we need to try many colourings of $G$ to find optimal colour-matching embeddings of $R$ into $G$.
Thus, while it is possible to enumerate all embeddings of each cycle and path when (or just after) running Algorithm~\ref{alg:embed-cycle}, doing so may be unnecessary and will actually instead introduce extra delay between the output of solutions to \countkStatOpt.
Instead, Algorithm~\ref{alg:embed-cycle} will return a representation of the optimal embeddings.
This representation contains two parts: firstly, the set $P_b$ contains the states that correspond to colour-preserving embeddings of a cycle $R_a$ that maximise the value of $\sum_{j=1}^a I(\psi(r^a_j), E(\psi(R^a))$.
The second part is all sets $P_v$ for $v\in V(G)$.
A simple backtracking algorithm will then be able to use these parts to reconstruct all embeddings $\psi$ that maximise $\sum_{j=1}^a I(\psi(r^a_j), E(\psi(R^a))$.

\begin{algorithm}[pt]
\caption{Embed cycle}\label{alg:embed-cycle}
\KwData{A vertex-coloured cycle $R^a$ on $|V(R^a)|$ distinct colours such that $r^a_i$ has colour $c_i$, and a (not necessarily proper) vertex-coloured graph $G$ with $\Delta(G) = 3$ on the same set of $|V(R^a)|$ colours.}
\KwResult{The set $P_b$ of all states that correspond to optimal colour-preserving embeddings of $R^a$ into $G$, as well as sets $P_v$ for $v\in V(G)$ of states corresponding to partial colour-preserving embeddings which may lead to the optimal colour-preserving embeddings.}
\ForEach{vertex $v$ of $G$}{
    Let $S(v) = \emptyset$\;
}
\ForEach{vertex $v$ of $G$ of colour $c_1$}{\label{alg:embedc-startv}
    \ForEach{neighbour $w$ of $v$ of colour $c_2$}{\label{alg:embedc-ne}
    \ForEach{neighbour $y$ of $v$ of colour $c_\ell$}{\label{alg:embedc-ne2}
            \If{$(v,w,v,y)\not\in S(w)$}{ \label{alg:embedc-start-ia}
                Add $(v,w, v,y)$ to $S(w)$\;
                Let $I_w'(v,w,v,y) = I(v,\{vw,vy\})$\;
            }
            \Else{
                \If{$I(v,\{vw,vy\}) > I_w'(v,w,v,y)$}{
                    Let $I_w'(v,w,v,y) = I(v,\{vw,vy\})$\;\label{alg:embedc-start-ib}
                }
            }
        }
    }
}
\ForEach{$i\in\{2,\ldots,\ell-1\}$}{\label{alg:embedc-midc}
    \ForEach{vertex $v$ of $G$ of colour $c_i$}{\label{alg:embedc-midv}
        \ForEach{state $(u,v,x,y)\in S(v)$}{\label{alg:embedc-mids}
            \ForEach{neighbour $w$ of $v$ of colour $c_{i+1}$}{\label{alg:embedc-midn}
                \If{$(v,w,x,y)\not\in S(w)$}{  \label{alg:embedc-end-ia}
                    Add $(v,w,x,y)$ to $S(w)$\;
                    Let $I_w'(v,w,x,y) = I'_v(u,v,x,y) + I(v,\{uv,vw\})$\;
                    Let $P_w(v,w,x,y) = \{(u,v,x,y)\}$\;
                }
                \Else{
                    \If{$I(v,\{vw,vy\}) + I'_v(u,v,x,y) > I_w'(v,w,x,y)$}{
                        Let $I_w'(v,w,x,y) = I(v,\{vw,vy\}) + I'_v(u,v,x,y)$\;\label{alg:embedc-end-ib}
                        Let $P_w(v,w,x,y) = \{(u,v,x,y)\}$\;
                    }
                    \ElseIf{$I(v,\{vw,vy\}) + I'_v(u,v,x,y) = I_w'(v,w,v,y)$}{
                        Let $P_w(v,w,x,y) = P_w(v,w,x,y) \cup \{(u,v,x,y)\}$\;
                    }
                }
            }
        }
    }
}
Let $I_b = -\infty$\;
Let $P_b = \emptyset$\;
\ForEach{vertex $v$ of $G$ of colour $c_\ell$}{\label{alg:embedc-endv}
    \ForEach{state $(u,v,x,y)\in S(v)$}{\label{alg:embedc-ends}
        \If{$v = y$}{  \label{alg:embedc-end-inc}
            \If{$I_v'(u,v,x,y) + I(v, \{uv,xy\}) > I_b$}{
                Let $I_b = I'_v(u,v,x,y) + I(v, \{uv,xy\})$\;\label{alg:embedc-end-upd}
                Let $P_b = \{(u,v,x,y)\}$\;\label{alg:embedc-end-setp}
            }
            \ElseIf{$I_v'(u,v,x,y) + I(v,\{uv,xy\}) = I_b$}{
                Let $P_b = P_b \cup \{(u,v,x,y)\}$\;\label{alg:embedc-end-addp}
            }
        }
    }
}
Return $P_b$ as well as $P_v$ for all $v\in G$\;
\end{algorithm}

We now bound the number of states, and show the correctness and running time of Algorithm~\ref{alg:embed-cycle}.

\begin{lemma}\label{lemma:fpt:embed-cycle:statecount}
For any vertex $v$, Algorithm~\ref{alg:embed-cycle} stores at most $3^k$ states at $S(v)$.
\end{lemma}
\begin{proof}
For a given vertex $v$ there are three possible choices for $u$ (as $G$ has maximum degree at most three), for each $u$ there are $3^{k-2}$ choices for $x$ as $x$ must be at distance at most $k-2$ from $u$ in a graph of maximum degree at most three, and for each $x$ there are at most three choices for $y$ as $y$ is a neighbour of $x$.
\end{proof}

Recall that we will represent a set of optimal embeddings by storing two pieces of data. The first is $P_b$, the set of states that actually correspond to the optimal embeddings (recall that these states don't by themselves store enough to recreate a complete embedding).
Secondly, we store the sets $P_v$ for $v\in V(G)$, which we can use in a backtracking algorithm in combination with $P_b$ to reconstruct the optimal embeddings.

\begin{lemma}\label{lemma:fpt:embed-cycle:correct}
    For a given coloured cycle $R^a$, Algorithm~\ref{alg:embed-cycle} correctly returns a representation of all embeddings that achieve the best improvement to neighbourhood discrepancy over all embeddings of $R^a$ onto $G$ such that the colours of the vertices of $R^a$ and $G$ agree.
\end{lemma}
\begin{proof}
We first show that the algorithm correctly determines the states at each vertex, and correctly calculates the values of $I'$ for each state and vertex.
We proceed inductively on $i$, and will iterate through the colours $c_i$, beginning with $i=2$ as no states are stored, or need to be stored, at any vertex of colour $c_1$.
Let $w$ be an arbitrary vertex in $V(G)$ with colour $c_2$.
Recall that $R^a_i$ denotes the induced subgraph of $R^a$ induced by the first $i$ vertices.
Thus $R^a_2$ has only two vertices, and so for a state $(v,w,v,y)$ there must be an embedding $\psi$ that corresponds to $(v,w,v,y)$ such that $\psi(v^r_1v^r_2) = vw$ and $v$ is a neighbour of $y$.
We see that the algorithm exactly considers all vertices of colour $c_2$ on Line~\ref{alg:embedc-startv}, and considers all pairs of edges $vw$ and $vy$ on Lines~\ref{alg:embedc-ne} and \ref{alg:embedc-ne2}, and that no other edges are considered when calculating states, so the correct states are determined for vertices of colour $c_2$.
Next, consider $I'_w(v,w,x,y)$ for an arbitrary state $(v,w,x,y)$ at $w$.
As $w$ has colour $c_2$, by definition we must have $v=x$. We also know that $y$ is a neighbour of $v$ and $y$ has colour $c_\ell$.
Let $\psi$ be an embedding such that $I(\psi(r^a_1), \{vw,wy\}) = I'_w(v,w,x,y)$ (i.e., $\psi$ is an embedding that achieves the best improvement over all embeddings that correspond to $(v,w,w,y)$).
Note that $\psi$ maps $r^a_1$ to $v$, $r^a_2$ to $w$, and $r^a_\ell$ to $y$, and those vertices are the only vertices involved when calculating $I(v, \{vw, wy\}) = I'_w(v,w,x,y)$.
Therefore when the algorithm considers all pairs of edges $vw$ and $vy$ on Lines~\ref{alg:embedc-ne} and \ref{alg:embedc-ne2}, it must also find an embedding that will map $r^a_1$ to $v$, $r^a_2$ to $w$, and $r^a_\ell$ to $y$, and no other edges are considered when calculating states, so the values of $I'_w(v,w,x,y)$ are correctly determined for vertices of colour $c_2$.

We now proceed inductively.
For $i\in\{2,\ldots,\ell-1\}$, for any state $s$ at a vertex $w$ of colour $c_{i+1}$ there must be an embedding $\psi$ of $R^a_{i+1}$ that corresponds to $s$.
We can then consider the states at vertex $\psi(r^a_i) = v$, where $v$ must have colour $c_i$ and be a neighbour of $w$.
In particular, as $\psi$ is an embedding of $R^a_{i+1}$, there must be a valid state $s'$ at $v$ such that $\psi$ corresponds to $s'$.
As Line~\ref{alg:embedc-midv} considers all such vertices $v$, Line~\ref{alg:embedc-mids} considers all states at $v$, and Line~\ref{alg:embedc-midn} considers all neighbours of $v$, it must be that the state corresponding to the embedding $\psi$ is added to $S(w)$.
Additionally, no other states are added to $S(w)$ so the correct states are determined.

Next we show that, given a state $(v,w,x,y)$ at vertex $w$ with colour $c_i$, there is no embedding that achieves a better improvement than $I'_w(v,w,x,y)$ as determined by Algorithm~\ref{alg:embed-cycle}.
To aid readability, in this paragraph we outline how this is shown, while the following paragraph will give the exact mathematical proof.
Recall that if $w$ has colour $c_i$, then $I'_w(v,w,x,y)$ is the best improvement on the first $i-1$ vertices, as the improvement at vertex $w$ depends on how the next vertex is embedded.
We will take an embedding $\psi$ that corresponds to $(v,w,x,y)$ and achieves an improvement of $I_w(v,w,x,y)$ (i.e.,~$\psi$ achieves the best improvement over all embeddings considered by Algorithm~\ref{alg:embed-cycle}), and then towards a contradiction assume that there exists some other embedding $\psi'$ that also corresponds to $(v,w,x,y)$, but achieves a strictly better improvement on the first $i-1$ vertices (which implies that $\psi'$ is not considered by Algorithm~\ref{alg:embed-cycle}).
We create the embedding $\psi''$ to be equal to $\psi'$ on the first $i-2$ vertices, and equal to $\psi$ on the remaining vertices.
Then by its construction $\psi''$ must be considered by Algorithm~\ref{alg:embed-cycle}, but attain a strictly better improvement than $\psi$, a contradiction.

We now give the technical details for the previous paragraph.
Let $\psi$ be an embedding such that $I'_w(v,w,x,y) = \sum_{j=1}^{i-1} I(\psi(r^a_j),E(\psi(R^a)))$ (i.e.~$\psi$ gives a maximal improvement on the first $i-1$ vertices for the state $(v,w,x,y)$ over all embeddings considered by Algorithm~\ref{alg:embed-cycle}), and
let $u$ be the vertex in $V(G)$ such that $\psi(r^a_{i-2}) = u$.
Then $\psi$ must correspond to a state $(u,v,x,y)$ at $v$.
We claim that $\psi$ must also obtain the maximum improvement over the first $i-2$ vertices over all embeddings that correspond to $(u,v,x,y)$.
Towards a contradiction, assume there is some $\psi'\neq \psi$ that also corresponds to $(u,v,x,y)$ such that 
\[
\sum_{j=1}^{i-2} I(\psi(r^a_j),E(\psi(R^a))) < \sum_{j=1}^{i-2} I(\psi'(r^a_j),E(\psi'(R^a)))
\]
(i.e.,~$\psi$ will not give a maximal improvement over the first $i-2$ vertices for the state $(u,v,x,y)$).
Consider next the embedding $\psi''$ where $\psi'(r^a_j) = \psi''(r^a_j)$ for $j\in\{1,\ldots, i-2\}$ and $\psi(r^a_j) = \psi''(r^a_j)$ for $j >= i-1$.
We note that $\psi''(r^a_{i-2}) = u$ and $\psi''(r^a_{i-1}) = v$, and as $(u,v,x,y)$ is a state at $v$ it must be that $uv$ is an edge.
This, combined with both $\psi$ and $\psi'$ being embeddings, ensures that $\psi'$ is an embedding.
Then 
\[
I(\psi(r^a_{i-1},\psi(E(R^a))) = I(\psi(r^a_{i-1},\psi''(E(R^a)))
\]
and we get 
\[
\sum_{j=1}^{i-2} I(\psi(r^a_j),E(\psi(R^a))) < \sum_{j=1}^{i-2} I(\psi'(r^a_j),E(\psi'(R^a)))
\]
which is a contradiction.
Thus there is no embedding corresponding to $(u,v,x,y)$ that achieves a better improvement to the neighbourhood discrepancy than $\psi$ over the first $i-2$ vertices.
Therefore $I'_w(v,w,x,y) = I'_v(u,v,x,y) + I(v, \{uv,vw\})$ is the maximum improvement attained by any embedding of the first $i-1$ vertices.
By induction we therefore know that Algorithm~\ref{alg:embed-cycle} correctly calculates the maximum improvement attainable at a given state over all embeddings.

We now know that states calculated at each vertex that is not coloured with $c_\ell$ are correct.
Lines~\ref{alg:embedc-endv} through \ref{alg:embedc-end-upd} then check each vertex $v$ of colour $c_\ell$, and then check each state $(u,v,x,y)$ at such a vertex to see if $y = v$.
As we have been preserving colours when mapping $R^a$ into $G$, this ensures that this is a colour-preserving embedding of the cycle $R^a$.
The maximum value of $I'_v(u,v,x,y) + I(v, \{uv, xy\})$, as calculated on Line~\ref{alg:embedc-end-upd} is thus tracked as $I_b$ and this is used to store the states that achieve this optimal value.

Lastly, we claim that we return a representation of all embeddings that achieve this maximum.
These are returned in $P_b$ and $P_v$ for $v\in G$.
By Lines~\ref{alg:embedc-end-setp} and \ref{alg:embedc-end-addp}, and the above arguments, for any $v \in G$ with colour $c_\ell$ and for any embedding that achieves the optimal score and maps $\ell$ to $v$, we store in $P_b$ all of the states of $v$ that represent this embedding.

For any state $(u,v,x,y)\in P_b$, we can rebuild all embeddings $\psi$ that achieve this state as follows.
We first see that $\psi(r^a_\ell) = y$, $\psi(r^a_0) = x$, and $\psi(r^a_{\ell-1}) = u$.
Then we can consider all elements of the set $P_v(u,v,x,y)$. Each element of this set leads to at least one embedding.
Starting with $i=\ell-1$, if $(u,v,x,y)$ is our current state then for each element $(t,u,x,y) \in P_v(u,v,x,y)$, we know there is at least one optimal, colour-preserving embedding with $\psi(r^a_{i-1}) = t$.
By repeating this process with decreasing values of $i$  and letting $(t,u,x,y)$ be the new current state until $i=2$, we can rebuild any embedding that achieves this optimal.
Recall that, at this point, we have a set of states for each vertex, and each such state contains the previous vertex, so the next state to consider and the previous vertex can both be accessed in constant time.
We note that this rebuilding process takes $O(\ell)$ time per embedding found.
\end{proof}

We note that Algorithm~\ref{alg:embed-cycle} can easily be extended to also find optimal embeddings of paths by using sentinel values (like $\emptyset$) instead of the first and last vertices of the current cycle in the definition of states (i.e., each state will ``look like'' $(u,v,\emptyset,\emptyset)$ where $u,v \in V(G)$), and excluding the check on Line~\ref{alg:embedc-end-inc}.
This will not increase the number of states stored at any vertex.

\begin{lemma}\label{lemma:fpt:embed-cycle:complexity}
    Algorithm~\ref{alg:embed-cycle} can be implemented to run in $O(3^kn)$ time.
\end{lemma}
\begin{proof}
A vertex is considered in one of the three loops on Lines \ref{alg:embedc-startv}, \ref{alg:embedc-midv}, or \ref{alg:embedc-endv}. Additionally, if a vertex is considered on Line~\ref{alg:embedc-midv}, it is considered in one iteration of the loop begun on Line~\ref{alg:embedc-midc}, as the vertex has one colour.
For the loop on Line~\ref{alg:embedc-startv}, we iterate through all pairs of neighbours.
As there are at most three neighbours, there are at most six such pairs.
For any other of the loops, we need to iterate through states at a vertex (of which there are at most $3^k$ by Lemma \ref{lemma:fpt:embed-cycle:statecount}), and neighbours of the correct colour (again, there may be at most three neighbours), giving the result.

Lastly, we highlight that for any individual vertex $v$ and set $F$, the value of $I(v, F)$ can be calculated in $O(\max d(v), |F|)$ time, which in Algorithm~\ref{alg:embed-cycle} is constant time.
\end{proof}

\begin{cor}\label{coro:fpt:embed-cycle:enum-complexity}
    \enumEmbedCycle can be solved with $f(k) \cdot n^{O(1)}$ precalculation time and $O(k)$ delay time when parameterised by $k$.
\end{cor}
\begin{proof}
The embeddings can be enumerated by first running Algorithm~\ref{alg:embed-cycle} and then backtracking through the vertices of $R^a$ to build the actual embeddings.
By Lemma~\ref{lemma:fpt:embed-cycle:complexity}, Algorithm~\ref{alg:embed-cycle} runs in $O(3^kn)$ time, and rebuilding an embedding from the $P_b$ and $P_w(u,v,x,y)$ for vertices $w$ and states $(u,v,x,y)$ takes $O(k)$ time, giving the desired result.
\end{proof}

We now describe how to combine all of these results.

\thmfptkdeletionenum*

\begin{proof}
In the precalculation we determine the maximum value of the score function and build the data required to output all solutions. 
We do this by first considering all potential options for the graph $R = G \setminus H$.
For each such $R$ we consider all potential degree sequences of $H = G \setminus \psi(R)$ for some embedding $\psi$, and then for each pair of $R$ and degree sequence, we determine the maximum value the score function can take, and store sufficient data to recreate all embeddings that achieve said maximum.
We iterate this over such graphs $R$ and degree sequences.
In doing so, we determine the maximum score function, and also have created all information required to recreate all embeddings that achieve said optimal score.
Once this optimal score is found, we then simply iterate through all graphs $R$ that achieve this optimal solution, all degree sequences of $G \setminus \psi(R)$ for an embedding $\psi$ that achieve this optimal solution, and all embeddings $\psi$ that achieve this optimal solution.

By Lemma~\ref{lemma:deletion:subgraphs} there are $O(2^k e^{\sqrt{k}})$ choices for $R$, and we will consider each in turn, so for now consider one choice of such a graph.
By Lemma~\ref{lemma:deletion:degreesequences}, for a given $R$ there are $O(k)$ choices for the degree sequence of $H = G\setminus \psi(R)$ for some embedding $\psi$.
Again we will deal with each in turn, so for now consider one choice of a degree sequence.
In particular, this fixes $h$, the number of vertices of degree one in $R$ that are mapped to vertices of degree three in $G$.
As $R$ can have at most $k$ components, and thus at most $k$ paths and at most $2k$ vertices of degree one, there are $O(\binom{2k}{h}) = O(2^{2k})$ ways to choose $h$ vertices of degree one in $R$, so additionally we fix this choice.
Let $g$ be a function mapping the vertices of $R$ to $\{2,3\}$ such that if $g(v^r_i) = b$ then $v^r_i$ must be mapped to a vertex in $G$ of degree $b$.

Note that when calculating 
\[
  \score(H, f) = \sum_{v\in V(G)} \ln \deg_H(v) - n \ln \left[ \sum_{v\in V(G)}\deg_H(v) \ND_H(v,f) \right],
\]
knowing the degree sequence of $H$
exactly determines the value of 
\[
  \sum_{v\in V(G)} \ln \deg_H(v).
\]
It therefore remains to minimise the value of \[
n \ln \left[ \sum_{v\in V(G)} \deg_H(v) \ND_H(v,f)\right].
\]

For this, we use colour coding to embed the paths and cycles of $R$ iteratively.
Colour-coding involves colouring the vertices of $R$ with each colouring in a set of colourings in turn, and solving the problem on each colouring in turn.
We later show the existence of a suitable set $\mathcal{C}$ of colourings, where $|\mathcal{C}|$ is logarithmic in $n$ but exponential in $k$.
To begin with, however, assume we have coloured the vertices of $R$ with $|V(R)|$ colours such that each vertex has a different colour.
We also colour the vertices of $G$ with the same $|V(R)|$ colours, noting that this colouring of $G$ need not be a proper colouring.

However, we must also ensure that any embedding we find also maps the correct number of vertices of degree one in $R$ to vertices of degree three in $G$.
Recall that $R=\bigcup_{a=1}^{m} R^a$ where each $R^a$ is connected (i.e.,~each $R^a$ is either a cycle or path).
Then, for a given component $R^a$ of $R$, we take a copy $G'$ of $G$, and for each colour $c_i$, and for each vertex $v$ of $G'$ with colour $c_i$, if $d_G(v) \neq F(v^r_i)$ (i.e., $v$ does not have our desired degree to match our degree sequence), then delete $v$ from $G'$.
This modification means that any colour-preserving embedding of $R^a$ into $G'$ that we find must exactly map each vertex $v^r_i$ of degree one of $R$ to a vertex of degree $F(v^r_i)$ in $G$, and therefore we can guarantee that we obtain the desired degree sequence.
Thus, for each component $R^a$, create the corresponding graph $G'$ and then run
Algorithm~\ref{alg:embed-cycle} on $(R^a, G')$.
As each vertex in $R$ has a different colour, we know that any two distinct components $R^a$ and $R^b$ have no vertices in common, so the best improvement to the neighbourhood discrepancy found for any embedding of $R$ can be calculated by summing together the best improvement to the neighbourhood discrepancy found for embedding $R^a$ over all connected components $R^a$ in $R$.

It then remains to determine how many different colourings of $G$ must be processed to guarantee that an optimal solution is found.
This guarantee is given if, for any subset of vertices $S\subseteq V(G)$ with $|S| \leq |V(R)| \leq 2k$, there is some colouring  that is processed such that all vertices in $S$ have different colours.
Such a set of colourings is equivalent to a $2k$-perfect family of hash functions.
From~\cite{NaorDerandomise}, there exists a set $\mathcal{C}$ of colourings of $G$ with $|\mathcal{C}| = 2^{O(k)} k^{O(1)} \log n$ such that for any set $S$ of $2k$ vertices in $G$, there is a colouring $C\in \mathcal{C}$ such that each of the $2k$ vertices in $S$ are differently coloured by $C$ (and by the same paper, such a set can also be constructed in $k^{O(k)} \poly(n)$ time).

We therefore have $O(2^k e^{\sqrt{k}})$ choices for $R$ and $O(k^k)$ choices for our function $g$. For each pair $(R, g)$ we need to use $(2k)^{O(k)} \log n$ colourings.
From~\cite{NaorDerandomise}, such a set of colourings can be found in $k^{O(k)} \poly(n)$ time, and the best improvement for each colouring can be found in $O(k3^kn)$ time by Lemma~\ref{lemma:fpt:embed-cycle:complexity} and the fact that $R$ has at most $k$ components.

All of the above choices are checked in the precalculation step, and we keep a list of all choices of colourings, degree sequences and graphs $R$ that achieve the optimal value in the obvious manner.

Then, once the precalculation has been completed, we need only output the subgraphs that achieve this optimal.
Each embedding can be generated in $O(k)$ time (as all the requisite data has already been produced).
As $G$ has maximum degree three, it also has at most $3n/2$ edges and so we can re-construct each optimal subgraph $H$ in $O(n)$ time.
Thus, there is a precalculation time of $2^{k(2\log k + O(1))} n \log n$ before the first subgraph is output, and further embeddings have a delay of $O(n)$.

Lastly, the postcalculation time must include checks that there are no more choices of embeddings, colourings, degree sequences, or graphs $R$ to make.
As choices of colourings, degree sequences or graphs $R$ are each completely calculated in the precalculation step, the lack of further choices for these can be determined in constant time by checking whether we have tried all choices that lead to optimal embeddings.
However, choices of embeddings are calculated during the output step from the $P_w(u,v,x,y)$ sets, and so the post-calculation step to know that no more embeddings are available may take $O(n)$ time, as required.
\end{proof}

If $k$ is fixed to some constant integer, then as $G$ has at most $3n$ edges, the number of subgraphs of $G$ that have $|E(G)|-k$ edges is $n^{O(k)}$, which is polynomial in $n$.
Therefore the number of solutions is, at worst, polynomial in $n$, meaning our algorithm is guaranteed to run in polynomial time.

We note that the decision-variant of \countkStatOpt where we only want to either find one optimal embedding, or even the optimal value of the score function, is in FPT when parameterised by $k$, the maximum number of edges that can be removed, by solving \countkStatOpt and returning the appropriate value when the first embedding is output.

\begin{theorem}\label{thm:fpt-k-deletion-one}
   For an integer $k\geq 1$, \kStatOpt can be solved on graphs with maximum degree three in time $2^{k(2 \log k + O(1))} n \log n$.
\end{theorem}

\section{Discussion and Conclusions}\label{sec:conclusion}
\StatOpt is a graph optimisation problem arising from spatial statistics with direct applications to epidemiology and social science that we show is intractable unless P = NP.
We also show that it is resistant to common techniques in graph algorithms, but is in XP when parameterised by both treewidth and maximum degree, and is fixed-parameter tractable when parameterised by the number of edges that can be removed and the maximum degree is limited to three.
Both of our algorithms are extended to not just solve the decision but also output all optimal subgraphs.
This results in two enumeration algorithms which respectively have XP (in treewidth and maximum degree) and FPT (in the maximum number of edges that can be removed) precalculation times, linear delays and linear postcalculation times.
However the question still remains as to whether \StatOpt itself is hard when the maximum degree of the input graph is bounded.
We also note as an interesting open problem whether \StatOpt admits efficient parameterised algorithms with respect to (combinations of) parameters other than the maximum degree.
Additionally, the original paper that introduced \StatOpt gives one heuristic for solving the problem, but leaves open any guarantee on the performance of this heuristic.
Thus the investigation of the performance of this heuristic, or indeed of any new approximation algorithms, form two other significant open problems for \StatOpt.

\section*{Acknowledgements}
All authors gratefully acknowledge funding from the Engineering and Physical Sciences Research Council (ESPRC) grant number EP/T004878/1 for this work, while Meeks was also supported by a Royal Society of Edinburgh Personal Research Fellowship (funded by the Scottish Government).

\bibliography{Refs}
\bibliographystyle{plain}
\end{document}